\documentclass[twocolumn]{aastex63}

\usepackage{color}
\usepackage{listings}
\definecolor{lbcolor}{rgb}{0.9,0.9,0.9}
\newcommand{\simprop}{\mathrel{\vcenter{
  \offinterlineskip\halign{\hfil$##$\cr
    \propto\cr\noalign{\kern2pt}\sim\cr\noalign{\kern-2pt}}}}}
    
\graphicspath{{./}{figures/}}

\begin{document}
\title{SCExAO/CHARIS Near-IR Integral Field Spectroscopy of the HD 15115 Debris Disk}
\correspondingauthor{Kellen Lawson}
\email{kellenlawson@gmail.com}
\author{Kellen Lawson}
\affiliation{Department of Physics and Astronomy, University of Oklahoma, Norman, OK}
\author{Thayne Currie}
\affiliation{Subaru Telescope, National Astronomical Observatory of Japan, 
650 North A`oh$\bar{o}$k$\bar{u}$ Place, Hilo, HI  96720, USA}
\affiliation{NASA-Ames Research Center, Moffett Blvd., Moffett Field, CA, USA}
\affiliation{Eureka Scientific, 2452 Delmer Street Suite 100, Oakland, CA, USA}
\author{John P. Wisniewski}
\affiliation{Department of Physics and Astronomy, University of Oklahoma, Norman, OK}
\author{Motohide Tamura}
\affil{Astrobiology Center of NINS, 2-21-1, Osawa, Mitaka, Tokyo, 181-8588, Japan}
\affiliation{Department of Astronomy, Graduate School of Science, The University of Tokyo, 7-3-1, Hongo, Bunkyo-ku, Tokyo, 113-0033, Japan}
\affiliation{National Astronomical Observatory of Japan, 2-21-2, Osawa, Mitaka, Tokyo 181-8588, Japan}
\author{Glenn Schneider}
\affil{Steward Observatory, The University of Arizona, Tucson, AZ 85721, USA}
\author{Jean-Charles Augereau}
\affil{Univ. Grenoble Alpes, CNRS, IPAG, 38000 Grenoble, France}
\author{Timothy D. Brandt}
\affiliation{Department of Physics, University of California, Santa Barbara, Santa Barbara, California, USA}
\author{Olivier Guyon}
\affiliation{Subaru Telescope, National Astronomical Observatory of Japan, 
650 North A`oh$\bar{o}$k$\bar{u}$ Place, Hilo, HI  96720, USA}
\affil{Steward Observatory, The University of Arizona, Tucson, AZ 85721, USA}
\affil{College of Optical Sciences, University of Arizona, Tucson, AZ 85721, USA}
\affil{Astrobiology Center of NINS, 2-21-1, Osawa, Mitaka, Tokyo, 181-8588, Japan}
\author{N. Jeremy Kasdin}
\affiliation{College of Arts and Sciences, University of San Francisco, San Francisco, CA, USA}
\affiliation{Department of Mechanical Engineering, Princeton University, Princeton, NJ, USA}
\author{Tyler D. Groff}
\affiliation{NASA-Goddard Space Flight Center, Greenbelt, MD, USA}
\author{Julien Lozi}
\affiliation{Subaru Telescope, National Astronomical Observatory of Japan, 
650 North A`oh$\bar{o}$k$\bar{u}$ Place, Hilo, HI  96720, USA}
\author{Jeffrey Chilcote}
\affiliation{Department of Physics, University of Notre Dame, South Bend, IN, USA}
\author{Klaus Hodapp}
\affiliation{Institute for Astronomy, University of Hawaii,
640 North A`oh$\bar{o}$k$\bar{u}$ Place, Hilo, HI 96720, USA}
\author{Nemanja Jovanovic}
\affiliation{Department of Astronomy, California Institute of Technology, 1200 East California Boulevard, Pasadena, CA 91125}
\author{Frantz Martinache}
\affiliation{Universit\'{e} C\^{o}te d'Azur, Observatoire de la C\^{o}te d'Azur, CNRS, Laboratoire Lagrange, France}
\author{Nour Skaf}
\affiliation{Subaru Telescope, National Astronomical Observatory of Japan, 
650 North A`oh$\bar{o}$k$\bar{u}$ Place, Hilo, HI  96720, USA}
\affil{LESIA, Observatoire de Paris, Université PSL, CNRS, Sorbonne Universit\'e, Universit\'e de Paris, 5 place Jules Janssen, 92195 Meudon, France}
	\affil{Department of Physics and Astronomy, University College London, London, United Kingdom}

\author{Eiji Akiyama}
\affiliation{Department of Engineering, Niigata Institute of Technology,
1719 Fujihashi, Kashiwazaki, 945-1195, Japan}

\author{Thomas Henning}
\affiliation{Max Planck Institute for Astronomy, Königstuhl 17, D-69117 Heidelberg, Germany}

\author{Gillian R. Knapp}
\affiliation{Department of Astrophysical Science, Princeton University, Peyton Hall, Ivy Lane, Princeton, NJ08544, USA}

\author{Jungmi Kwon}
\affiliation{Department of Astronomy, Graduate School of Science, The University of Tokyo, 7-3-1, Hongo, Bunkyo-ku, Tokyo, 113-0033, Japan}

\author{Satoshi Mayama}
\affiliation{The Graduate University for Advanced Studies, SOKENDAI,
Shonan Village, Hayama, Miura, Kanagawa 240-0193, Japan}

\author{Michael W. McElwain}
\affiliation{NASA-Goddard Space Flight Center, Greenbelt, MD, USA}

\author{Michael L. Sitko}
\affiliation{Space Science Institute, 475 Walnut Street, Suite 205, Boulder, CO 80301, USA}

\author{Ruben Asensio-Torres}
\affiliation{Department of Astronomy, Stockholm University, AlbaNova University Center, SE-106 91, Stockholm, Sweden}

\author{Taichi Uyama}
\affiliation{Infrared Processing and Analysis Center, California Institute of Technology, 1200 E. California Boulevard, Pasadena, CA 91125, USA}
\affiliation{NASA Exoplanet Science Institute, Pasadena, CA 91125, USA}
\affiliation{National Astronomical Observatory of Japan, 2-21-2, Osawa, Mitaka, Tokyo 181-8588, Japan}

\author{Kevin Wagner}
\affiliation{Steward Observatory, The University of Arizona, Tucson, AZ 85721, USA}

\accepted{July 31, 2020}
\submitjournal{AJ}

\begin{abstract}
We present new, near-infrared ($1.1 - 2.4$ $\mu m$) high-contrast imaging of the debris disk around HD 15115 with the Subaru Coronagraphic Extreme Adaptive Optics system (SCExAO) coupled with the Coronagraphic High Angular Resolution Imaging Spectrograph (CHARIS). SCExAO/CHARIS resolves the disk down to $\rho \sim 0\farcs2$ ($\rm{r_{proj}} \sim 10$ $\rm{au}$), a factor of $\sim 3-5$ smaller than previous recent studies. We derive a disk position angle of $\rm{PA}$ $\sim 279\fdg4 - 280\fdg5$ and an inclination of $\rm{i}$ $\sim 85\fdg3 - 86.2\fdg$. While recent SPHERE/IRDIS imagery of the system could suggest a significantly misaligned two ring disk geometry, CHARIS imagery does not reveal conclusive evidence for this hypothesis.  Moreover, optimizing models of both one and two ring geometries using differential evolution, we find that a single ring having a Hong-like scattering phase function matches the data equally well within the CHARIS field of view ($\rho \lesssim 1\arcsec$). The disk's asymmetry, well-evidenced at larger separations, is also recovered; the west side of the disk appears on average around 0.4 magnitudes brighter across the CHARIS bandpass between 0\farcs25 and 1\arcsec.
Comparing STIS/50CCD optical photometry ($2000-10500$ \AA) with CHARIS NIR photometry, we find a red (STIS/50CCD$-$CHARIS broadband) color for both sides of the disk throughout the $0\farcs4 - 1\arcsec$ region of overlap, in contrast to the blue color reported at similar wavelengths for regions exterior to $\sim 2\arcsec$. Further, this color may suggest a smaller minimum grain size than previously estimated at larger separations.
Finally, we provide constraints on planetary companions, and discuss possible mechanisms for the observed inner disk flux asymmetry and color.
\end{abstract}


\section{Introduction} \label{sec:intro}

Gas-poor, dusty debris disks around stars are key laboratories for studying planetary system structure and the late stages of their formation \citep{Wyatt2008, Hughes2018}. Scattered light imagery of debris disks around young stars clarifies the disks' structures and can identify the signatures of sculpting planets and in-situ formation and erosion of icy Kuiper belt objects \citep{Kalas2005,Kenyon2008}. Further, these studies enable analysis of the composition and scattering properties of the material within the disks, potentially providing reference points for the evolution of the Kuiper belt \citep{Currie2015b}. High-contrast imaging produced using so-called ``extreme adaptive-optics'' (exAO) facilities (e.g. SPHERE \citep{Beuzit2019}, GPI \citep{Macintosh2015}, and SCExAO \citep{Jovanovic2015,Lozi2018,Currie2019b}) provides the opportunity to study these systems to smaller inner working angles than was possible with conventional AO. This enables the assessment of the disks' scattering phase functions at previously inaccessible angles, as well as placing more significant constraints on the presence of embedded planets.

The debris disk around HD 15115, an F2V star at a distance of $49.0 \pm 0.1$ pc \citep{Gaia2018} and an estimated age of $\sim 10-100$ Myr \citep[e.g.][]{Moor2006, Rhee2007, Gagne2018}, could be a particularly good target for studying planetary system structure and the results of the initial formation stages. Discovery optical scattered-light imagery from the Hubble Space Telescope (HST) Advanced Camera for Surveys (ACS) reveals a highly inclined disk with an ``extreme" east-west length asymmetry in the HST/ACS F606W bandpass ($\lambda_{pivot} = 5886$ \AA, FWHM $= 2325$ \AA), resolving the eastern extent out to $\sim 7\arcsec$ while the western extent is recovered to the edge of the field of view at $12\farcs38$ \citep{Kalas2007}. Their measurements of the disk's surface brightness on either side show an approximately symmetric brightness at $2\farcs0$, with the west becoming brighter than the east at larger separations ($\Delta m \sim 1$ at $6\arcsec$). Follow-up $H$-band Keck/NIRC2 adaptive-optics (AO) imagery resolved the disk at $\rho \sim 1\arcsec - 3\farcs3$ and revealed a blue (F606W-H) color on both sides and a brightness asymmetry beyond 2\arcsec{}. The combination of the reported color and its highly inclined orientation led to HD 15115's disk being informally referred to as ``the Blue Needle''. 
Follow-up observations from ground-based AO and space expanded the wavelength range over which HD 15115's disk is resolved and further clarified its properties at separations beyond 1\arcsec{}. HST/NICMOS 1.1 $\micron$ data revealed evidence of more complicated color gradients and a wavelength dependence for the disk's asymmetry and an angular separation dependence for its colors \citep{Debes2008}.

Subsequent studies found evidence of a bow-like shape in the disk at $\sim$ 1\arcsec{}--2\arcsec{} \citep{Rodigas2012,Mazoyer2014,Sai2015}, consistent with a ring-like disk at $\sim$ 90 au.
Using archival near-infrared imaging from the Gemini Observatory, \citet{Mazoyer2014} conclude that, while the system's ring is asymmetrical in brightness, the geometry of the ring itself is symmetric about the parent star. \citet{Schneider2014} reported HST Space Telescope Imaging Spectrograph (STIS) data which significantly improved upon the visible light photometry and morphology of the disk. These data showed that the bowing and asymmetry in visible wavelengths continue down to 0\farcs{}4. Additionally, this HST/STIS imaging revealed a previously unseen morphological bifrucation on the east side of the outer disk \citet{Schneider2014}.

Recent results suggest the possible existence of multiple debris ring components. \citet{Engler2019} reported the first extreme AO observations of HD 15115, consisting of VLT/SPHERE total intensity data in J and H band, and polarized intensity data in J band. They recover the disk over stellocentric separations of $\rho \sim 1\farcs0 - 5\farcs5$ and suggest, from peaks in their polarized intensity profiles, the possibility of a distinct non-coplanar inner disk having a fiducial radius of $\sim 1\farcs3$. Attempting to investigate this using their total intensity imagery, they are unable to reveal conclusive evidence regarding the disk's geometry. \citet{MacGregor2019} reported 1.3 mm Atacama Large Millimeter/submillimeter Array (ALMA) observations of the system with a synthesized beam size of $0\farcs58 \times 0\farcs55$. From these observations, evidence exists for either a distinct inner ring, with radius $\sim 0\farcs95$, or a significant gap in the canonical disk at a separation of $1\farcs2$. Notably, however, they report a lack of evidence in their data to support the misalignment of the inner disk hypothesized by \citet{Engler2019}. Additionally, \citet{MacGregor2019} found an absence of the east-west brightness asymmetry typically reported in previous NIR and optical imagery (e.g. \citealt{Kalas2007,Mazoyer2014}). They suggest that the large-grain dust population probed by ALMA was unaffected by the mechanism responsible for the asymmetry reported by other studies over similar separations. As perturbations from planetary mass companions are often used to explain disk asymmetries and more complex, multi-ringed disk geometries \citep[e.g.][]{MacGregor2019}, the details of these occurrences in the HD 15115 disk are significant.  To better clarify the presence or absence of additional ring components and brightness asymmetries over a wide wavelength range, high-contrast imaging data interior to 1\arcsec{}, matching the coverage of STIS data from \citet{Schneider2014}, are needed.

In this work, we report new near-infrared scattered light imagery of the HD 15115 system using the Subaru Coronagraphic Extreme Adaptive Optics (SCExAO) system and the Coronagraphic High Angular Resolution Imaging Spectrograph (CHARIS) integral field spectrograph in broadband (spanning near-infrared J, H, and K bands, $1.13-2.39$ $\micron$) mode \citep{Groff2016}. This imagery provides a view of the disk to separations a factor of $\sim 3-5$ smaller than previous recent studies ($\rho \sim 0\farcs2$). We conduct analysis of the disk's color in NIR and optical wavelengths by combining CHARIS IFS data with prior HST STIS imagery. Through both spine tracing and forward modeling, we investigate the details of the system's geometry and offer constraints for the presence of additional rings or planet companions within CHARIS's $2\arcsec \times 2\arcsec$ field of view.

\section{Data}\label{sec:data}
\subsection{Observations}
HD 15115 was observed on 2017 August 30 and  2017 September 07 using the Subaru Telescope’s SCExAO paired with the CHARIS integral field spectrograph operating in low-resolution (R $\sim 20$), broadband (1.13–2.39 $\mu m$) mode, and utilizing SCExAO's Lyot coronagraph with 217 mas diameter occulting spot. CHARIS has a nominal pixel scale of 0\farcs0164 $\rm{pixel}^{-1}$, which has been revised to 0\farcs0162 $\rm{pixel}^{-1}$ \citep{Currie2018}. Both sets of data were collected in \textit{angular differential imaging} mode (ADI; \citealt{Marois2006}), achieving total parallactic angle rotations of $\Delta \rm{PA}=$ $76^{\circ}$, and $56^{\circ}$ with total integration times of $t_{int} =$ 81 and 55 minutes respectively. Each set is made up of 80 individual exposures, with August 30 images having exposure times of 60.48 seconds and September 07 images having exposure times of 41.3 seconds. Sky frames were obtained for both data sets.

For the September 7 data, the conditions were good, with the ``slow" (long coherence time) seeing having a full-width at half-maximum (FWHM) in $V$ band of $\theta_{\rm V}$ $\sim$ 0\farcs5. SCExAO's real-time telemetry data estimated $H$-band Strehl ratios of $\sim$ 80\%. The conditions for the August 30 data were comparable. No telemetry data were recorded for the August observation, but the point spread function (PSF) quality appeared slightly superior by-eye.

Additionally, we make use of HST/STIS analysis-quality (AQ) imagery (STIS/50CCD, $2000-10500$ \AA, $\lambda_{pivot} = 5752$ \AA), originally reported and analyzed in \citet{Schneider2014}, to better explore the colors of HD 15115's disk (see Section \ref{sec:color}).

\subsection{CHARIS Data Reduction}\label{sec:reduc}
CHARIS data were extracted from raw CHARIS reads using the CHARIS Data Reduction Pipeline \citep{Brandt2017}. Extracted data take the form of image cubes with dimensions $(N_\lambda, N_x, N_y) = (22,201,201)$ (i.e. $201 \times 201$ pixel images for each of 22 wavelength channels). Subsequent basic image processing -- e.g. sky subtraction, image registration, spectrophotometric calibration --  was carried out as in \citet{Currie2011,Currie2018}.

PSF subtraction was performed by application of both the \textit{Karhunen-Lo\`{e}ve Image Projection} (KLIP; \citealt{Soummer2012}) and the \textit{Adaptive, Locally Optimized Combination of Images} (A-LOCI; \citealt{Currie2012,Currie2015}) algorithms independently.

We performed PSF subtraction with settings geared towards the detection of a) the HD 15115 debris disk and b) companions plausibly responsible for sculpting the disk. Table \ref{tab:psfsub_settings} lists our parameter choices for each reduction, with the motivations for these choices summarized hereafter.

\textbf{Disk Detection} -- 
The HD 15115 debris disk is oriented nearly edge-on in the plane of the sky \citep[e.g.][]{Kalas2007, Mazoyer2014}.  To detect the disk, we performed PSF subtraction exploiting ADI only, not SDI. For A-LOCI, tuning the geometry of the optimization and subtraction regions -- i.e. their relative azimuthal and radial widths -- is essential toward the recovery of disk flux.
Combining the minimum rotation gap with azimuthally elongated regions allowed LOCI coefficients to be computed with less perturbation by the radially extended disk flux while still producing a strong reconstruction of the speckle noise.   For KLIP, we performed PSF subtraction in full annuli.   To limit self-subtraction of the disk,  we imposed a minimum rotation gap of $\delta \sim 1.25 - 1.50$ $\lambda / D$  when selecting suitable reference frames for both A-LOCI and KLIP reductions.

\textbf{Companion Detection} -- To achieve deeper contrast limits needed to detect faint planets, we used a combination of ADI and then SDI (on the ADI residuals) using A-LOCI following \citet{Currie2018}.   For the ADI component, optimization regions were a factor of 20 smaller (50 PSF footprints) and the rotation gap was reduced to $\delta$ = 0.5--0.75.  For a given section of the science image, up to the 50 most correlated sections from the reference image library were used to build a reference PSF (with the number of available images depending on the portion of the 80 exposures satisfying the minimum rotation gap requirement for the section).   For the SDI component, the optimization zone covers an annular region with the same depth $\Delta$~$r_{\rm sub}$ as the subtraction zone but the smaller annular wedge-shaped subtraction zone is masked.

    \begin{deluxetable*}{@{\extracolsep{8pt}}cccccccccc}
    \tablewidth{0pt}
    \tablecaption{PSF Subtraction Algorithm Settings}
    \tablehead{\colhead{} & \colhead{} &\multicolumn{4}{c}{A-LOCI} & \multicolumn{4}{c}{KLIP} \\
    \cline{3-6}
    \cline{7-10}
    \colhead{Data} & \colhead{Parameter Tuning} &\colhead{$g$} & \colhead{$N_A$} & \colhead{$\delta_{\rm{FWHM}}$}  & \colhead{$\Delta r_{\rm{sub}}$} & \colhead{$N_{\rm{PCA}}$} & \colhead{$N_{\rm{zones}}$} & \colhead{$\delta_{\rm{FWHM}}$} &  \colhead{$\Delta r_{\rm{sub}}$}}
    \startdata
    Aug 30 & disk & 0.1 & 1000 & 1.50 & 4 & 5 & 1 & 1.25 & 4 \\
    Sep 07 & disk & 0.1 & 1000 & 1.50 & 4 & 5 & 1 & 1.25 & 4 \\
    Aug 30$^{a}$ & companion & 1.0 & 50 & 0.5, 1 & 10\\
    Sep 07$^{a}$ & companion & 1.0 & 50 & 0.75, 1 & 10
    \enddata
    \tablecomments{Algorithm settings for A-LOCI and KLIP utilized for PSF subtraction of each of the three data sets. `g' refers to the aspect ratio of the optimization regions, with $g < 1$ producing azimuthally elongated sections and $g > 1$ producing radially elongated sections. `$N_A$' refers to the area of optimization regions in units of PSF cores. `$\delta_{\rm{FWHM}}$' indicates the minimum rotation gap in units of PSF FWHM (for both A-LOCI and KLIP). `$\Delta r_{\rm{sub}}$' gives the radial size of subtraction regions in units of pixels (for both A-LOCI and KLIP). `$N_{\rm{PCA}}$' indicates the number of principal components utilized in construction of the model PSF. `$N_{\rm{zones}}$' is the number of subsections into which each KLIP optimization annulus was divided (with a value of 1 corresponding to full annuli). In all A-LOCI reductions, we also truncated the covariance matrix to zero out (normalized) singular values smaller than 1.25$\times$10$^{-6}$ and constructed a reference PSF from only the 50 most correlated images. $a)$ The two entries for the rotation gap refer to the ADI rotation gap and the SDI radial movement gap.   }\label{tab:psfsub_settings}
    \end{deluxetable*}

\subsection{Results}\label{sec:results}

\begin{figure*}
\includegraphics[width=\textwidth]{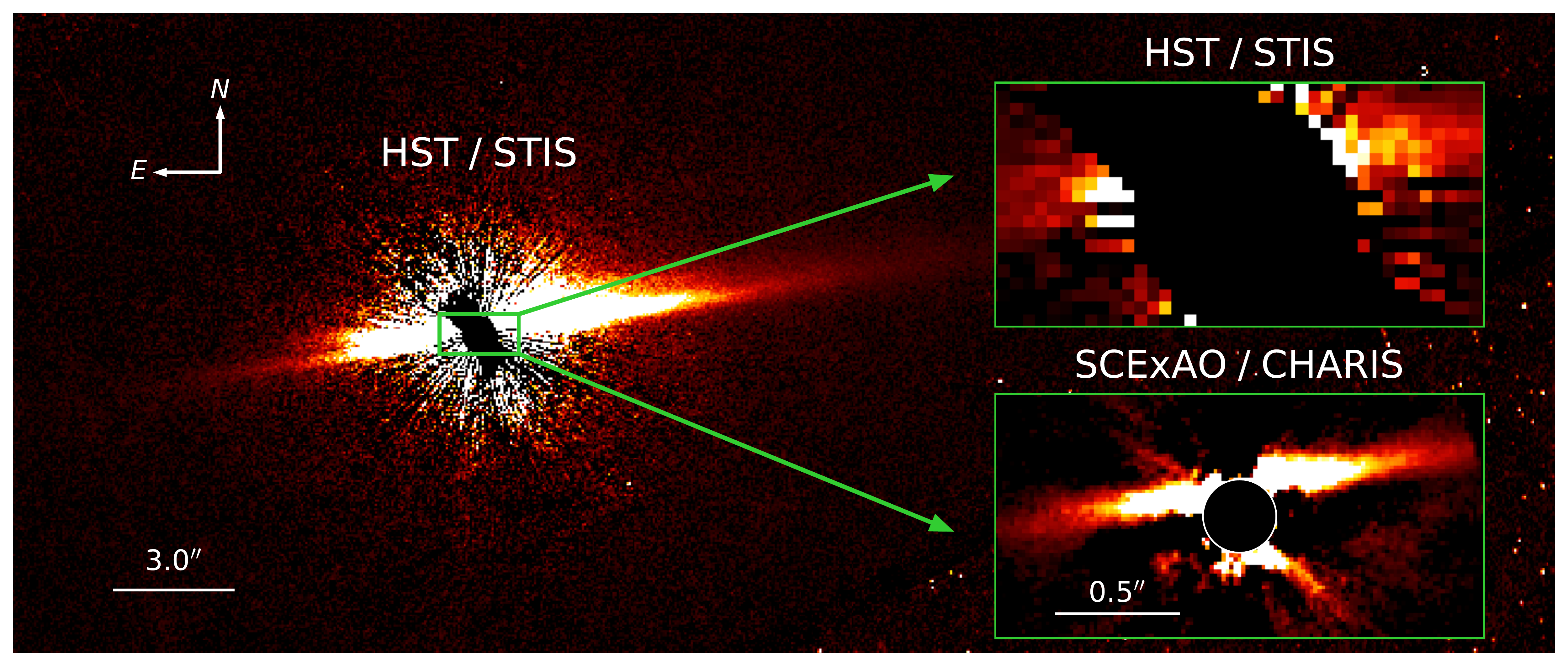}
\caption{HST/STIS imagery of HD 15115 (originally reported in \citet{Schneider2014}) with lower and upper inset images corresponding to CHARIS imagery and the same HST/STIS imagery scaled to the CHARIS field of view. CHARIS imagery presented is the average of the results for A-LOCI reductions of August 30 and September 07 data using settings for disk detection (see Table \ref{tab:psfsub_settings}). The CHARIS image's central mask has a radius of 0\farcs{}15. The STIS, STIS inset, and CHARIS images have the same orientation and are depicted with linear display stretches spanning $0-0.02$ $mJy$ $arcsec^{-2}$, $0-4.0$ $mJy$ $arcsec^{-2}$, and $0-5.21$ $mJy$ $arcsec^{-2}$ respectively.
\label{fig:charis_inset}
}
\end{figure*}

\begin{figure*}
\includegraphics[width=\textwidth]{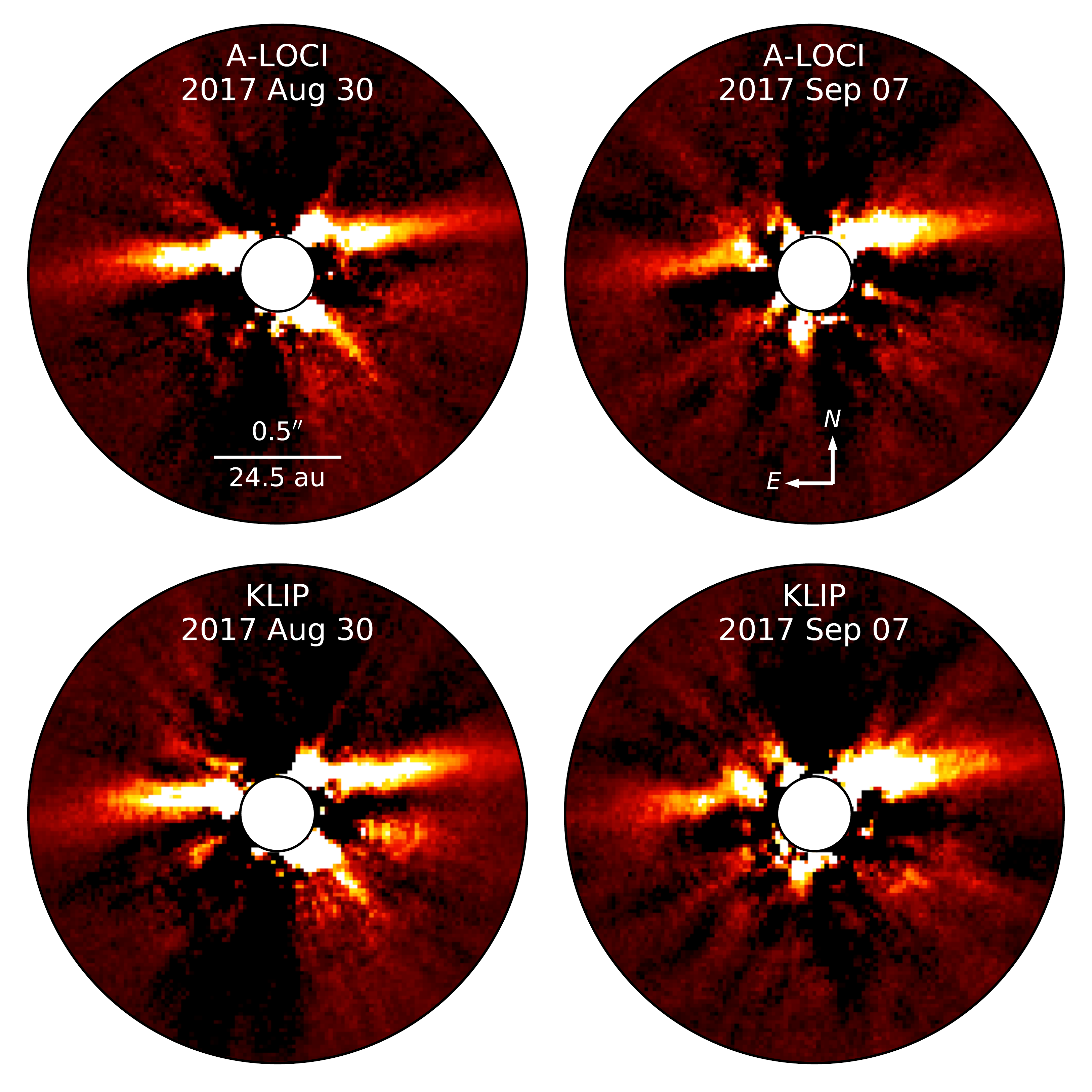}
\caption{
Wavelength-collapsed results for August 30 and September 7 observations following PSF subtraction using either A-LOCI or KLIP techniques with settings for disk detection (see Section \ref{sec:reduc}). The central mask in each subplot has a radius of 0\farcs{}15, and the image has a linear display stretch spanning $-0.61-6.10$ $mJy$ $arcsec^{-2}$. In August 30 products, a plainly visible flux enhancement appears just beyond the inner software mask to the southwest, but is not evident in September 07 data. This feature is likely residual speckle noise (likewise for the similar feature to the southeast in September 07 imagery).
\label{fig:reducs}
}
\end{figure*}

Both PSF subtraction techniques yield strong detections of the disk to $\rho \sim$ 0\farcs{}15 $-$ 0\farcs{}25 in CHARIS data (Figures \ref{fig:charis_inset}, \ref{fig:reducs}), improving upon the $0\farcs4$ angular separation achieved with optical HST/STIS data \citet{Schneider2014}. CHARIS data mark a substantial improvement over previous ground-based, near-IR scattered light imaging of the disk, with conventional AO data limited to $\rho$ $\gtrsim$ 1\arcsec{} and extreme AO imaging from \citet{Engler2019} detecting the disk exterior to $\rho \sim$ 0\farcs75 - 1\arcsec. This improvement is owed in part to the significant field rotation achieved, with $\Delta \rm{PA}=$ $76^{\circ}$ and $\Delta \rm{PA}=$ $56^{\circ}$ for our two sets of observations, versus e.g. $\Delta \rm{PA}=$ $23^{\circ}$ for the data from \citet{Engler2019}.

\begin{figure*}
\includegraphics[width=\textwidth]{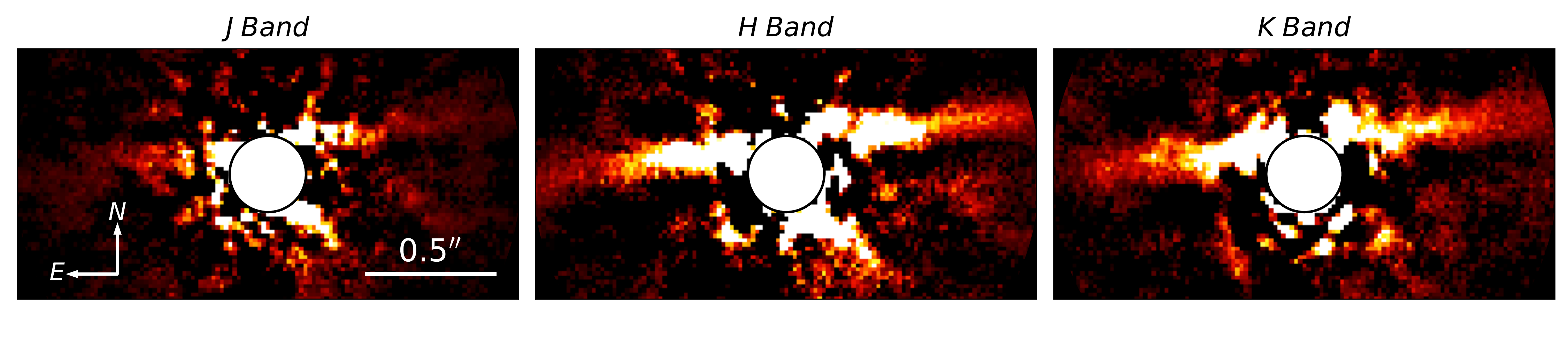}
\caption{
A-LOCI PSF subtracted imagery for August 30 using settings for disk detection, with wavelength channels combined to produce images comparable to J (channels $1-5$, $1.16-1.33$ $\micron$), H (channels $8-14$, $1.47-1.80$ $\micron$) and K (channels $16-21$, $1.93-2.29$ $\micron$) bands. The central mask in each subplot has a radius of 0\farcs{}15. Images are displayed with linear stretches spanning $0-24.13$ $mJy$ $arcsec^{-2}$ (J-band) or $0-4.39$ $mJy$ $arcsec^{-2}$ (H-band and K-band).
\label{fig:jhkbands}
}
\end{figure*}

\begin{figure*}
\includegraphics[width=\textwidth]{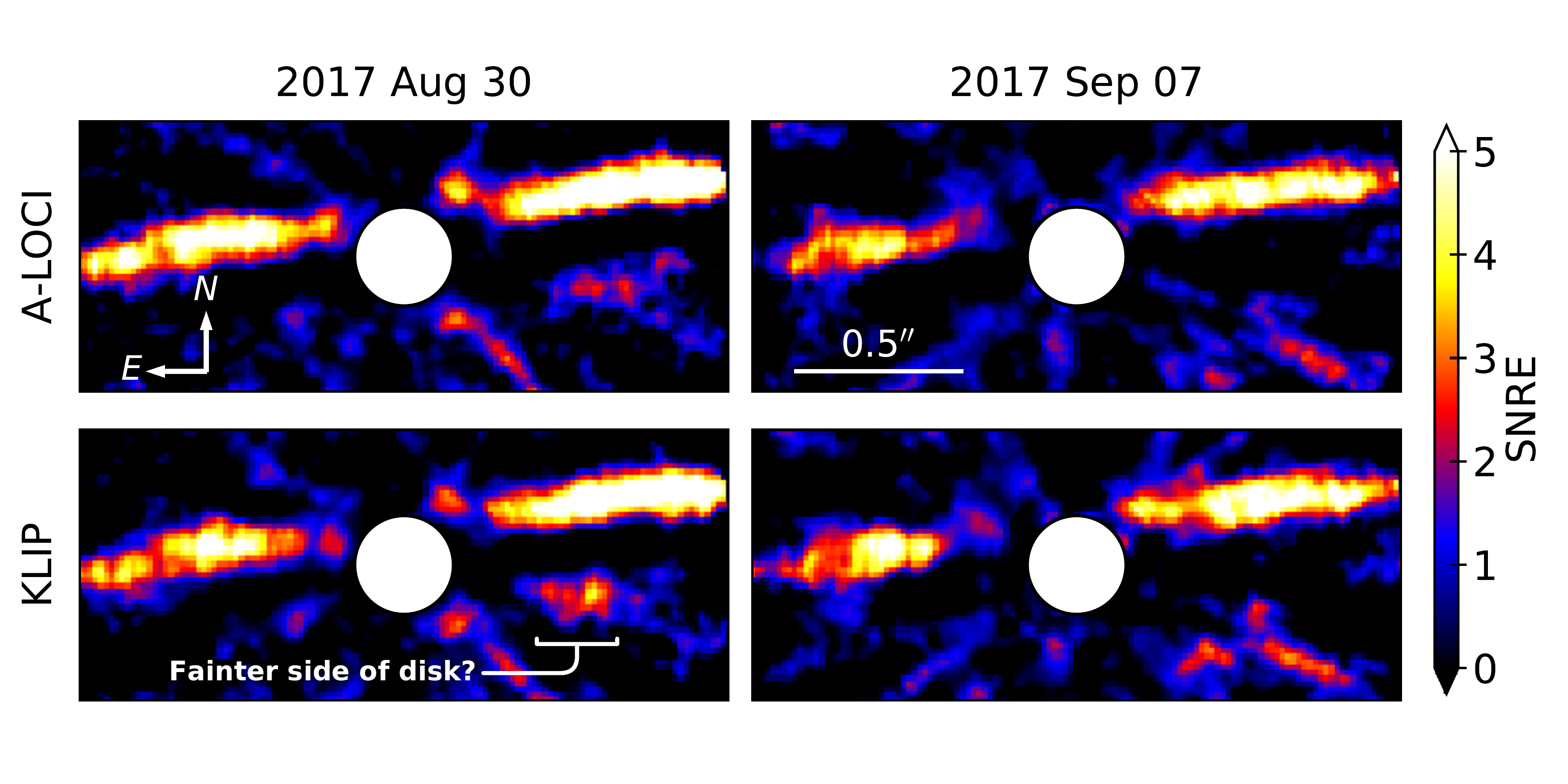}
\caption{
Maps of Signal-to-Noise per resolution element for August 30 and September 7 data following PSF subtraction using A-LOCI or KLIP techniques with settings for disk detection (see Section \ref{sec:reduc}). The central mask in each subplot has a radius of 0\farcs{}15. Possible signal from the disk's fainter side is indicated in the subplot for the KLIP reduction of August 30 data, peaking at a SNRE of $\sim 4$. This feature is also visible in the A-LOCI reduction of the same data, albeit at a lower SNRE ($\sim 2.5$).
\label{fig:snrmaps}
}
\end{figure*}

\begin{figure*}
\includegraphics[width=\textwidth]{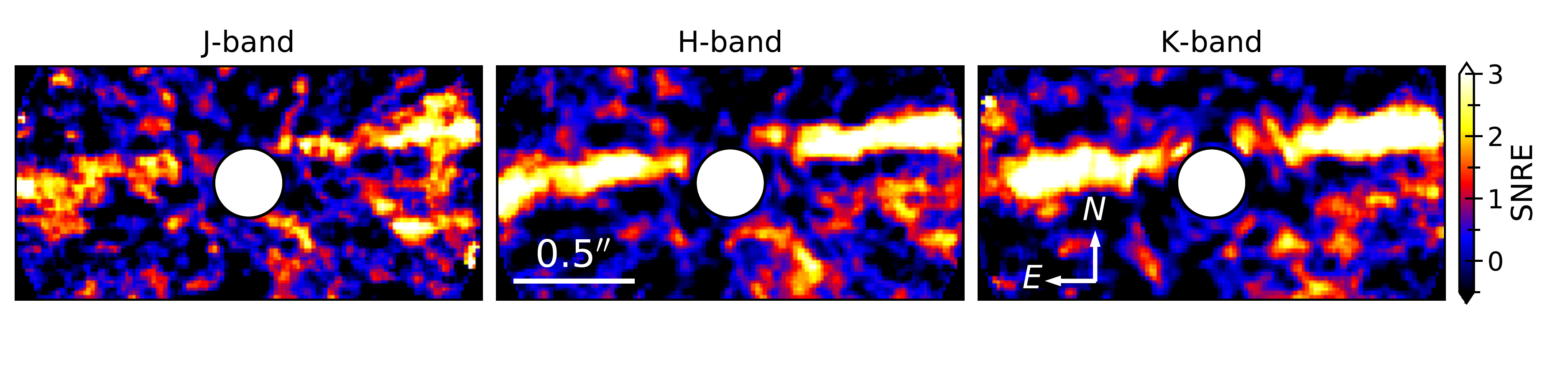}
\caption{
Maps of Signal-to-Noise per resolution element for J, H and K bands from A-LOCI PSF subtraction of August 30 data (see Section \ref{sec:reduc}). The central mask in each subplot has a radius of 0\farcs{}15.
\label{fig:jhksnrmaps}
}
\end{figure*}

The quality of the detection varies from $J$ band, where the disk detection is contaminated by residual speckles, to $H$ and $K$ band where the images are free of strong residuals exterior to 0\farcs{}25 (Figure \ref{fig:jhkbands}).  The detection in the broadband images is strongest in the Aug 30 data (especially for the A-LOCI reduction), with a signal-to-noise per resolution element (SNRE) along the trace of the disk of $\sim$ 5--7 for most regions exterior to 0\farcs{}25\footnote{In the SNR calculation, we use a software mask to reduce the amount of disk signal included in the noise estimation.   This approach increases the finite-element correction penalty \citep{Mawet2014}, so the gain in SNR is small.} (Figure \ref{fig:snrmaps}). In $H$ and $K$ bands, the disk detection is strong, generally achieving SNRE $\sim 3-5$ along the disk, and peaking around $5.6$ to the west; the $J$ band detection of the disk is considerably weaker (though still definitive), with SNRE $\sim 2-3$ over the same regions and peaking around $4$ in the west (see Figure \ref{fig:jhksnrmaps}).

CHARIS imagery reveals a strongly asymmetrically scattering disk whose maximum intensity is unambiguously offset (with a projected semi-minor axis of $\sim 0\farcs12$) from the system's major axis throughout the $\sim 2\arcsec \times 2\arcsec$ field of view. This indicates a view of the system entirely inside the bow-like feature originally described by \citet{Rodigas2012}. For the assumption of preferential forward scattering \citep{Hughes2018}, the brighter ($\sim \rm{northern}$) side of the disk observed clearly in our data would be presumed as the near side. Signal to the west and slightly south of the center in Figures \ref{fig:reducs} \& \ref{fig:snrmaps} (annotated in the latter) may constitute marginal detections of the disk's dimmer (presumably far) side, which has been recovered in previous ground-based imagery (e.g. \citealt{Mazoyer2014, Engler2019}).

\section{Disk Morphology} \label{sec:disk_spine}
\begin{figure*}
\includegraphics[width=\textwidth]{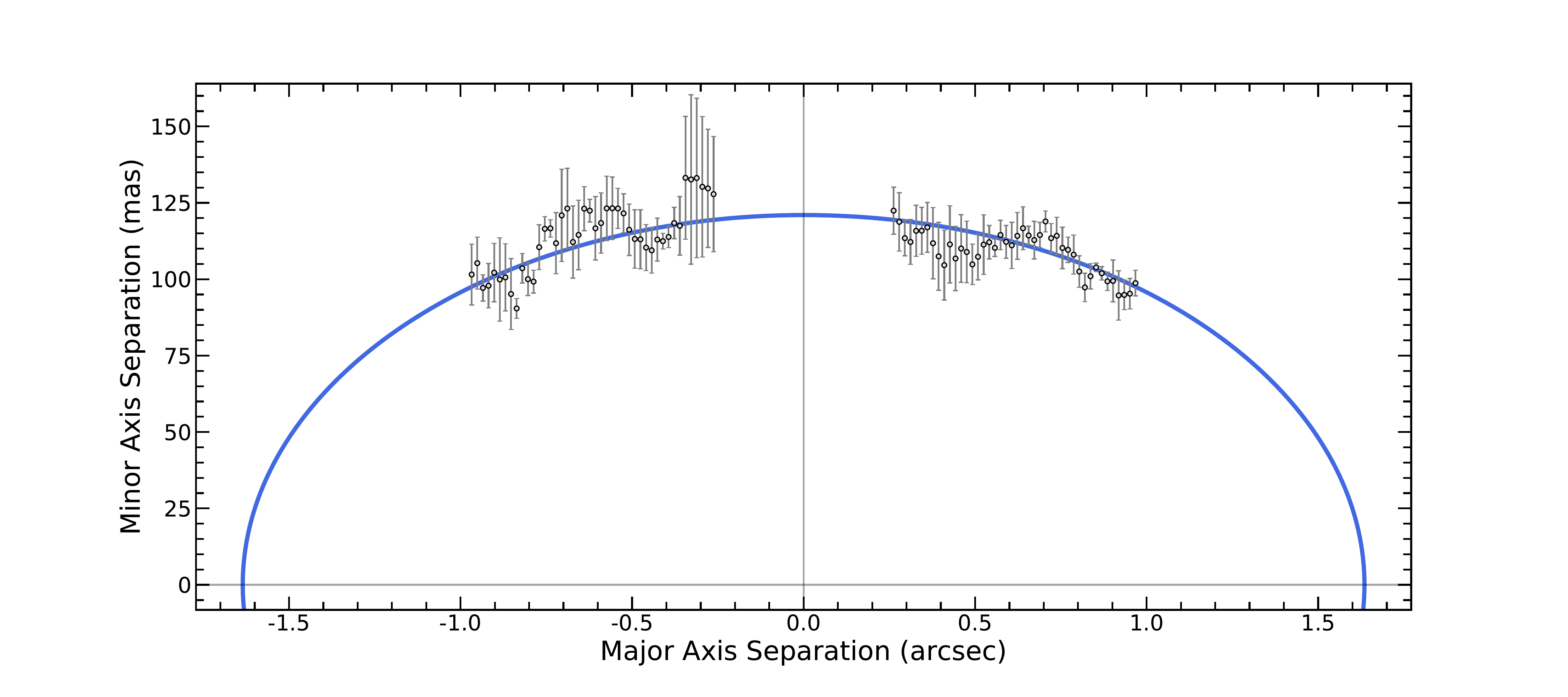}
\caption{
Stellocentric separation of disk spine fits along the major and minor axes. The spine positions are used for ellipse fitting and evaluating disk geometry. Points with errors correspond to the weighted average position from all four utilized CHARIS broadband images (see Section \ref{sec:disk_spine}). The blue arc depicts the best-fit ellipse solution for the data shown, and provides the disk major and minor axes against which the spine positions are plotted here.
\label{fig:spine}
}
\end{figure*}

To make estimates of the disk's geometric parameters, we seek to identify the position of peak brightness along the disk's bright ($\sim$ northern) edge for each of the wavelength-collapsed (CHARIS broadband) final images resulting from both A-LOCI and KLIP reductions of the August 30 and September 07 data (4 images in total). We begin by rotating the image based on the PA reported in \citet{Engler2019} (PA = $278\fdg9 \pm 0\fdg1$, which is then adjusted to account for the CHARIS PA correction discussed below) to orient the disk's major axis along the x-axis\footnote{The stated PA is assumed here only for the purpose of this initial rotation, which is carried out to simplify fitting of the spine across the narrowest part of the disk, where the peak will appear sharpest. This additionally allows meaningful measurements of the disk's projected FWHM for later use (see Section \ref{sec:color}). Rotating the images based on the 1-$\sigma$ upper and lower limits from \citet{Engler2019} instead (279\fdg0 and 278\fdg8 respectively) changes the eventual PA measurement from this procedure by $\sim 0\fdg01$, likely largely as a result of the rotation interpolation itself.}. For each x-direction integer pixel position, we identify the approximate y position of the brightness peak by taking the median of the locations of the brightest $5\%$ of pixels in that column. To determine the precise peak position, we then fit a Lorentzian profile to the array of flux values within 15 pixels ($\sim$0\farcs25) of the approximate peak, seeking the y position of the profile's peak (and taking the standard error from this fit as the uncertainty). For these fits, each flux value is weighted by the inverse of the corresponding noise levels from noise maps  (described in Section \ref{sec:forward_modeling}). The y-position of the spine as a function of x-position is computed this way for each of the four images being analyzed. The spine profiles for the four images are then combined by taking the weighted average of the four values at each x-position as the nominal average spine y-position, with uncertainty corresponding to the standard error for a weighted average\footnote{For the $j^{th}$ image's $i^{th}$ x-axis spine position, call the corresponding y-axis spine position $y_{ij}$ with associated fit y-position uncertainty $\sigma_{ij}$. The standard error for the corresponding weighted average, $\mu_i$, with weight $w_{ij} = 1 \; / \; \sigma^2_{ij}$, is then:
$$\sigma'_{i} = \sqrt{\frac{\sum_{j=1}^4 (y_{ij}-\mu_i)^2  \cdot w_{ij}}{\sum_{j=1}^4 w_{ij}}}$$
}. The resulting average spine profile is then fit with an ellipse (which is centered on the star), described by projected semi-major axis (\rm{a}), nominal position angle ($\rm{PA}_0$, which accounts for the initial image rotation applied previously) and inclination (\rm{i}). Our best fit (over the region $0\farcs25 \leq \rho \leq 1\farcs0$) is achieved for an ellipse with parameters: $\rm{a}=80.1 \pm 3.3$ au, $\rm{PA_0} = 277\fdg82 \pm 0\fdg05$, and $\rm{i} = 85\fdg76 \pm 0\fdg22$. The nominal position angle measurement is then corrected for the CHARIS PA offset of $-2\fdg20 \pm 0\fdg27$ \footnote{Hereafter, any values of PA presented (e.g. in the case of disk modeling in Section \ref{sec:disc_modeling}) are already corrected for this PA offset.} \citep{Currie2018}. This results in a final measurement of $\rm{PA} = 280\fdg02 \pm 0\fdg27$. The average spine positions and the best fit ellipse are visualized in Figure \ref{fig:spine}.

The position angle measured by this methodology falls above recent measurements made by \citet{MacGregor2019} ($\rm{PA} = 278 \pm 1 \degr$) and \citet{Engler2019} ($\rm{PA} = 278\fdg9 \pm 0\fdg1$). The measured inclination is consistent with the values of both works (\citet{MacGregor2019} measured $\rm{i} = 86\fdg3 \pm 0\fdg4$ and \citet{Engler2019} measured $\rm{i} = 85\fdg8 \pm 0\fdg7$). Evaluating $\chi^2_\nu$ for the spine parameters from \citet{Engler2019} with our data suggests that the difference we measure is significant, with these parameters giving $\chi^2_\nu = 36.6$ versus our best-fit of $\chi^2_\nu = 1.1$. Given that our measurements of the disk are made in the region of $\rho \sim$ 0\farcs2--1\farcs0, it is possible that we are measuring overlapping signal of the canonical outer ring and the inner ring proposed in both works. If the disk profile observed is the result of an architecture featuring a distinct non-coplanar inner ring (a possibility suggested by \citet{Engler2019}), we should expect the fit values to be skewed somewhere between those of the inner and outer component. The difference in measured PA could also be explained by an imprecise calibration for either instrument (or both). However, the CHARIS PA calibration utilized was performed using data collected only a day after our September observations \citep{Currie2018}: it should provide a reasonable assessment of the PA calibration for our data.  A reevaluation of the CHARIS north PA and pixel scale using additional data obtained at additional epochs reaffirm these results (T. Currie 2020, in prep.). 

Attenuation of disk flux during PSF subtraction can also have an effect on the measured position of the spine, and thus on the derived parameters as well. However, we note that carrying out the aforementioned measurements on attenuated models with known PA from our forward-modeling procedure (see Section \ref{sec:forward_modeling}) indicates that this effect is small. e.g. for our best-fit one ring model with true $\rm{PA} = 279\fdg8$, we measure $\rm{PA} = 279\fdg7 \pm 0\fdg3$.

\section{Modeling the Debris Disk of HD 15115}
\subsection{Disk Forward Modeling}\label{sec:forward_modeling}
We implement a strategy of forward-modeling synthetic disks, as described in \citet{Currie2018,Currie2019}, to investigate the details of HD 15115's debris disk. In this approach, coefficients (for A-LOCI) or Karhunen-Lo\`{e}ve modes (for KLIP) retained from the science data reduction are applied to image cubes containing only the signal of a model disk which has been rotated to reproduce the array of observed position angles and convolved with the instrumental point-spread function.

For this procedure, we consider all three sources of flux annealing described by \citet{Pueyo2016}: over-subtraction (speckle noise being subtracted from disk signal), direct self-subtraction (resulting from inclusion of disk signal in basis vectors), and indirect self-subtraction (resulting from perturbation of basis vectors by disk signal). For our disk reductions, indirect self-subtraction is expected to be the smallest of the three terms. For our KLIP reductions, we retained a small number of KL modes compared to the total number of reference images available.  For A-LOCI, we adopted a large optimization area ($N_{\rm A}$ = 1000 PSF cores) that is azimuthally elongated, in contrast to the nearly edge-on debris disk.  For both reductions, we adopted a large rotation gap of $\delta$ = 1.25--1.5 PSF cores. Thus, oversubtraction and self-subtraction likely dominate over indirect self-subtraction.

Once processed this way, the result for a given model can be compared to the result for science data to assess the relative strength of the model. The procedure for calculating $\chi^2_\nu$ that we implement is as previously described in \citet{Goebel2018}, but is briefly summarized here. First, each value in the model image, $f_{model}$, and the science image, $f_{obs}$, is replaced with the sum of values within a FWHM-sized aperture. Following this, a finite element corrected noise map is computed from the science image as described in \citet{Currie2011}, additionally utilizing a software mask as described above in Section \ref{sec:results}. Then, the model image is rescaled to minimize the inverse-variance-weighted residuals with the science image in a region of interest. The difference of the scaled model and the science image, weighted by the noise map, is squared to create a $\chi^2$ map. This map is then binned to the size of the instrumental PSF. Finally, the reduced $\chi^2$ metric is computed from this as $\chi^2_\nu = \chi^2 / \nu = \nu^{-1} \sum_i^N (f_{i,obs} - f_{i,model})^2 / \sigma_i^2$, where the degree of freedom, $\nu$, is given by the difference between the number of bins in the binned optimization region, $N$, and the number of free parameters in the model, $M$ \citep{Thalmann2013}.

The region of interest considered is a rectangular box of un-binned dimensions 120 pixels $\times$ 30 pixels ($\sim 2\farcs0 \times 0\farcs5 $) centered on the star and oriented to fall along the disk's approximate major axis. The region interior to 6 pixels ($\sim 0\farcs1$) is excluded. For $\rho \lesssim$ 6 pixels, not-a-number, or NaN, values begin to appear when no reference frames can meet the minimum rotation gap requirement during PSF subtraction. The region of interest described is overlaid as a white rectangle in Figures \ref{fig:onecomp_model} and \ref{fig:twocomp_model}). 

We delineate models which are acceptably consistent with our observations as in \citet{Thalmann2013}, i.e. those having $\chi^2_{\nu} \leq \chi^2_{\nu,min} + \sqrt{2 / \nu}$.

\subsection{Model Debris Disks}\label{sec:models}
The forward modeling procedure outlined above was applied to model debris disks generated using a version of the GRaTeR software \citep{Augereau1999}. The formalism and assumptions of the models are detailed in \citet{Augereau1999}, but summarized briefly hereafter. The models assume an optically thin disk with a radial dust grain distribution described by a smooth combination of two power laws and with a vertical distribution described by an exponential function. For simplicity, and in line with the analysis of \citet{Engler2019}, we set the vertical exponential distribution to be Gaussian in shape ($\gamma = 2$), and restrict the flaring of the disk to be linear ($\beta = 1$). To describe the angular distribution of scattered light, it is common to adopt the Henyey-Greenstein (HG) phase function \citep{Henyey1941}. The HG phase function is parameterized by a single variable, the asymmetry parameter $g$, defined as the average of the cosine of the scattering angle, weighted by the (normalised) phase function, over all directions. However, as noted in \citet{Hughes2018}, this formalism is not physically motivated and may introduce misleading results. Moreover, a simple HG phase function fails at reproducing the surface brightness profile we observe for HD 15115's disk (see Appendix \ref{app:phasefuncs}). \citet{Hong1985} implements a linear combination of three HG phase functions to describe the observed angular distribution of scattered light for zodiacal dust. To limit model freedom and avoid non-physical solutions, we adopted the phase function of \citet{Hong1985} with the asymmetry parameters and corresponding weights identified therein: $g_1 = 0.7$, $g_2 = -0.2$, and $g_3 = -0.81$, with weights $w_1 = 0.665$, $w_2 = 0.330$, and $w_3 = 0.005$. Though allowing the asymmetry parameters and weights to vary during exploration of disk models may improve the eventual result, it would also massively increase the size and complexity of the parameter space. As early testing showed that the empirically derived parameters of \citet{Hong1985} reproduced our observed disk surface brightness quite closely, we chose to adopt them as-is to allow a more thorough exploration of the remaining disk parameters.
See Appendix \ref{app:phasefuncs} for comparisons of scattering phase functions with our data, including simple HG phase functions using commonly reported asymmetry parameters for HD 15115's disk.

Model geometries investigated fall under two archetypes. The first is a single ring model, defined by 6 parameters:
\begin{enumerate}
\item $R_0$, the radius of peak grain density in au
\item $\alpha_{in}$, the power law index describing the change in radial density interior to $R_0$
\item $\alpha_{out}$, the power law index describing the change in radial density exterior to $R_0$
\item $\frac{H_0}{R_0}$, the ratio of disk scale height at $R_0$ to $R_0$
\item \textrm{PA}, the position angle of the disk in degrees
\item \textrm{i}, the inclination of the disk in degrees
\end{enumerate}

The second is a two ring model, taken to be the linear superposition of two single ring models\footnote{e.g. by coadding the synthetic images for the individual models, as in \citet{Boccaletti2019}. This assumes that the rings are sufficiently optically thin that single-scattering dominates over multiple-scattering.} plus an additional parameter: $F_{max,2} / F_{max,1}$, the ratio of the peak flux of ring 2 to that of ring 1. This results in a model described by 13 parameters (allowing inclination and position angle to differ between the inner and outer disks). To better explore the parameter space of the inner ring, we reduce these to 7 parameters by setting the well-studied outer ring's parameters to approximately match the ones identified by prior studies of the disk (e.g. \citealt{Engler2019}): $R_{0,1}$ = 96 au, $\alpha_{in,1}$ = 2, $\alpha_{out,1}$ = -3, $\frac{H_{0,1}}{R_{0,1}}$ = 0.03, \textrm{PA}$_1$ = 278\fdg9, and \textrm{i}$_1$ = 86\fdg0.

For the purpose of $\chi^2_\nu$ calculation, the overall scaling factor applied to the model (see Section \ref{sec:forward_modeling}) is considered to be an additional free parameter for the model. This results in $M = 7$ for the one ring model and $M = 8$ for the two ring model.

\subsection{Model Optimization Using Differential Evolution}\label{sec:model_opt}
Though HD 15115's disk has been studied extensively in the region beyond 1\farcs0, the small separations observed with CHARIS provide a look at the disk in the $0\farcs2 - 1\farcs0$ regime, where the parameters of the posited inner ring could potentially be studied in much greater detail. However, given the nearly edge-on orientation of the system, a parameter space with significant model degeneracies and multiple local minima is possible. A broad but detailed search of the parameter spaces outlined in Section \ref{sec:models} is necessary to offer a meaningful assessment of any such degeneracies and to ensure that a unique and globally optimum solution is identified. A grid search for the 7 parameter two ring model quickly reaches an intractably large size; e.g., a coarse grid examining only 5 values of each parameter would require 78125 models be propagated through the time consuming forward modeling procedure (typically $\sim$ minutes per model). While Markov-Chain Monte-Carlo (MCMC) techniques are commonly used for similar purposes (e.g \citealt{MacGregor2019}), they can become trapped in local minima and their results can be dependant on the initialization. Moreover, MCMC exploration typically requires a number of model evaluations that is effectively unapproachable for ADI forward modeling procedures comparable to ours (e.g. the MCMC procedure of \citealt{MacGregor2019} evaluates $\sim 10^6$ models).

Instead, we make use of the differential evolution algorithm (DE, \citealt{Storn1997}) to explore possible solutions for the parameters of each model. DE requires no initial assumptions about a solution, beyond boundaries within which to explore, and is capable of efficiently probing large, correlated parameter spaces by quickly evolving a population of trial solutions away from regions that offer inferior solutions and allowing population members to move between local minima. Though DE has not seen widespread use in the optimization of scattered light disk models, it has been used elsewhere in the study of astrophysics with noteworthy efficacy, e.g. to explore optimal capture trajectories for Jovian orbiters by the European Space Agency \citep{Labroqure2014}, to identify edge-on galaxies with abundances of extraplanar dust \citep{Shinn2018}, or to search for flaring stars in sparsely sampled time-series data \citep{Lawson2019}. While DE does not enable the robust determination of parameter likelihood distributions in the same manner as MCMC, it is extremely effective at quickly reaching a global solution with little to no tuning of algorithm control variables or specific experience with the algorithm itself, and with relatively few function evaluations needed (e.g. see comparison benchmarks in \citealt{Storn1997}).  Additionally, the relative simplicity of DE makes it trivial to add into existing frameworks (the C-style psuedo-code for the algorithm presented in \citealt{Storn1997} requires only 19 lines of code). For groups currently exploring model parameters using grid searches: in all but the rarest cases, DE will tend to identify a superior final model while evaluating many fewer models overall. For groups that might be interested in adopting this technique, we include a simple Python implementation of differential evolution, whose procedure is outlined below, in Appendix \ref{app:de_code}.

In the differential evolution procedure, we initialize a population of $N_{pop}$ random model parameter sets, restricted to fall between boundaries set for each parameter. For each type of model, $N_{pop}$ is set to be 10 times the number of free parameters\footnote{\citealt{Storn1997} suggest 5--10 population members per free parameter, though larger values are often used in recent implementations as well, e.g. the default value of 15 per free parameter in the implementation from the Python package SciPy \citep{Virtanen2020}.} (60 for the single ring model, and 70 for the two ring model). The initial model population is run through the forward modeling routine to evaluate the fitness of the models ($\chi^2_{\nu}$). Following this, a mutation ($v_i$) for each member of the population ($x_i$) is created by adding a scaled difference of the parameters of two distinct, random population members ($x_j$, $x_k$) to the parameters of the current best solution ($x_{best}$): $v_i = x_{best} + m (x_j - x_k)$ (this ``strategy'' is called  ``best/1/bin'' by the notation of \citealt{Storn1997}). For our purposes, the value of the mutation constant, $m$, is randomly selected in the range [0.5, 1.0] for every generation\footnote{\citealt{Storn1997} introduce a single mutation constant and suggest a value of 0.5. The ``dithered'' mutation constant implemented here is adopted from the default setting in the SciPy module for Python, where the authors suggest that a dithered mutation constant will typically speed convergence substantially \citep{Virtanen2020}.}. From a given mutation, a trial replacement ($u_i$) is formed by setting $u_i = x_i$ and allowing probability $P = 0.7$\footnote{A crossover probability, P, anywhere from 0.1 to 0.9 is recommended in \citealt{Storn1997}. P=0.7 is adopted from the default value in the SciPy module for Python.} for each parameter value in $u_i$ to be replaced by the corresponding value from $v_i$. The fitness of the set of $N_{pop}$ trial replacements is then evaluated using the same forward modeling routine. Finally, each member of the population, $x_i$, is replaced by its corresponding trial vector, $u_i$, if the fitness of the trial vector is superior to that of the population member. The procedure of creating and evaluating trial replacements for the population is repeated until the population becomes stagnant or converges to a single solution. 

Fitness is evaluated at each stage by computing the combined $\chi^2_\nu$ for both A-LOCI reductions as: $$\chi^2_{\nu} = \frac{\chi^2_1 + \chi^2_2}{N_1 + N_2 - M}= \frac{1}{\nu} \sum_{i=1}^2 \chi^2_i\; ,$$ where subscripts 1 and 2 correspond to the values for A-LOCI reductions of August 30 and September 07 data. Once the DE procedure is completed, the subset of models meeting the threshold $\chi^2_{\nu} \leq \chi^2_{\nu,min} + \sqrt{2 / \nu}$ are then propagated through the forward modeling procedure for the KLIP reductions. The final best model is taken from among these as the one which minimizes the combined $\chi^2_{\nu}$ for all four reductions: $$\chi^2_{\nu,\rm{tot}} = \frac{\sum_{i=1}^4 \chi^2_i}{(\sum_{i=1}^4 N_i)-M} = \frac{1}{\nu} \sum_{i=1}^4 \chi^2_i\; ,$$ with $i=3$ and $i=4$ indicating the two KLIP reductions. While it may be preferable to evaluate each set of trial models for all four reductions, this procedure cuts the total model optimization time nearly in half by assuming that the overall best model will be contained within the `acceptable' fitness bounds of the first two reductions. See column 2 of Table \ref{tab:onecomp_results} and Table \ref{tab:twocomp_results} for the bounds adopted for each parameter.

\subsection{Model Results}\label{sec:model_results}
\begin{figure*}
\includegraphics[width=\textwidth]{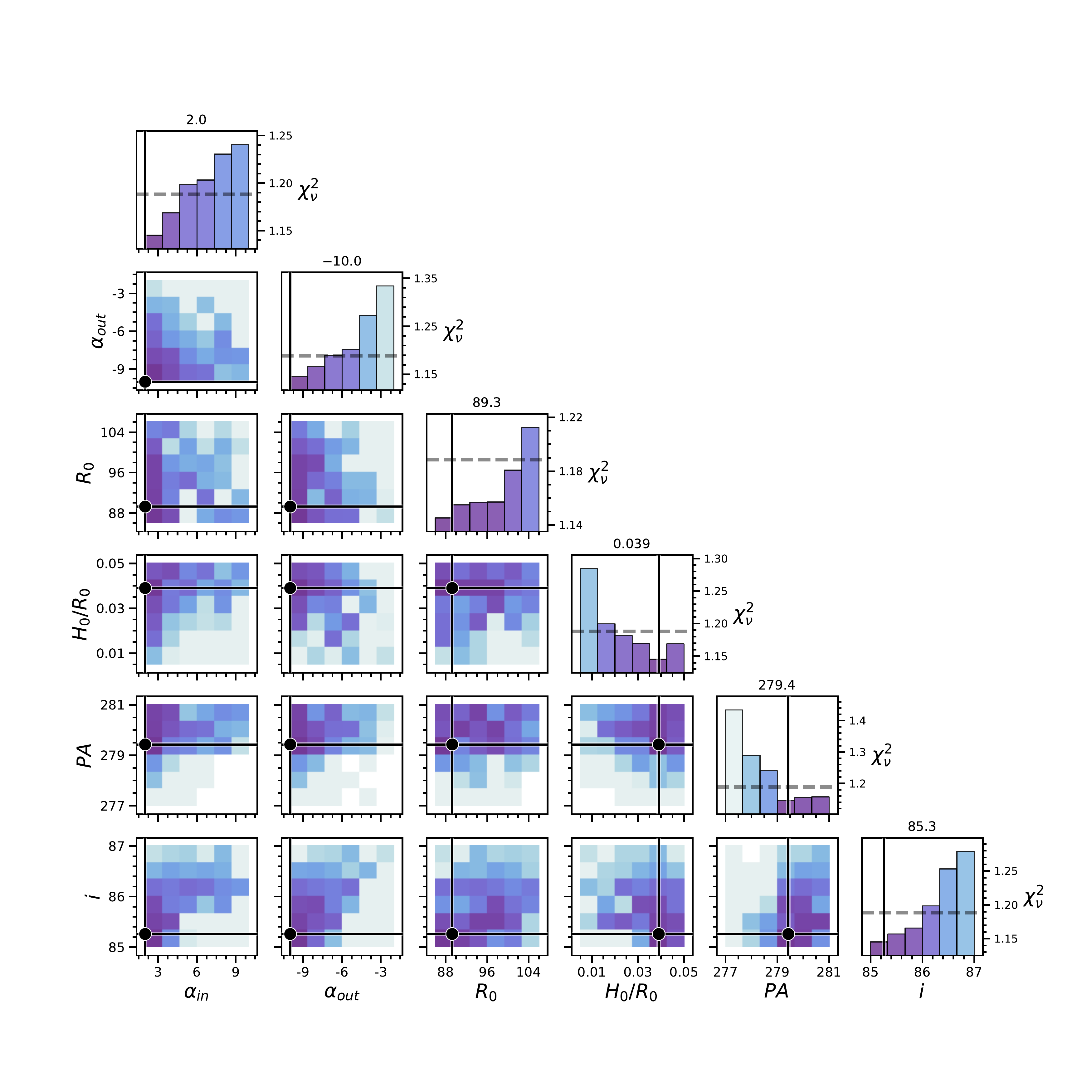}
\caption{A variation of the corner plot for optimization of the single ring model to the A-LOCI reductions of HD 15115's August 30 2017 and September 07 2017 data. Each off-diagonal plot visualizes solutions as a function of two of the parameters, with each bin colored according to the quality of the best fit achieved with values of the two parameters in that range (and any values of the other parameters). Darker bins indicate smaller values of $\chi^2_\nu$, where $\chi^2_\nu$ refers to the combined metric for the A-LOCI reductions of both data sets (see Section \ref{sec:model_opt}). Diagonal elements provide a one-dimensional view of each of the parameters, indicating the lowest $\chi^2_\nu$ value (y-axis) achieved for the binned range of the given parameter (x-axis). For each parameter, the area within the bounds provided in Table \ref{tab:onecomp_results} is divided into 6 equally sized bins. The best-fit solution values are indicated by black crosshairs, and their values given above the corresponding one-dimensional subplot. The threshold fitness for acceptable solutions with A-LOCI reductions ($\chi^2_{\nu} \leq \chi^2_{\nu,min} + \sqrt{2 / \nu}$) is indicated by a horizontal grey dashed line in each diagonal plot; bars which end below this threshold line resulted in models meeting the acceptable fitness criteria for some values of the other parameters.
\label{fig:onecomp_corner}
}
\end{figure*}

\textbf{One Ring Models} -- 
The differential evolution procedure for single ring models yields an optimal combined fit for the two A-LOCI reductions of $\chi^2_\nu$ = 1.145. A visualization of the full sample of models explored using differential evolution is provided in Figure \ref{fig:onecomp_corner}. Some parameters ($\alpha_{in}$ and $\alpha_{out}$) have converged to the boundaries. While this could indicate that the bounds are too restrictive, we note that the adopted boundaries include all values of these parameters explored by other recent studies of HD 15115 utilizing GRaTeR (e.g. \citealt{Mazoyer2014} use models with $-6 \leq \alpha_{out} \leq -4$ and $2 \leq \alpha_{in} \leq 10$, and \citealt{Engler2019} use models with $-8 \leq \alpha_{out} \leq -2$ and $2 \leq \alpha_{in} \leq 10$). This result is discussed in more detail in Section \ref{sec:disc_modeling}. However, we note here that these particular parameters are unimportant for our overarching conclusions.

From the 660 models evaluated, 13 resulted in acceptable values of $\chi^2_\nu$ ($\chi^2_\nu \leq 1.188$, for $\nu = 1077$ with the two A-LOCI reductions included). After forward modeling this subset for the remaining reductions, the final best model results in $\chi^2_{\nu,\rm{tot}} = 1.166$ with a revised acceptable limit of $\chi^2_{\nu,\rm{tot}} \leq 1.196$ (for $\nu = 2161$ when all four reductions are included). The parameters for the best overall model and ranges of acceptable parameters are included in Table \ref{tab:onecomp_results}. It should \textit{not} be assumed that an acceptable solution can be produced for any arbitrary combination of parameter values falling within the acceptable ranges. A given value included in the acceptable range indicates that the value, paired with specific values of the other parameters, produces a model meeting the given threshold for acceptability.

\begin{figure*}
\includegraphics[width=\textwidth]{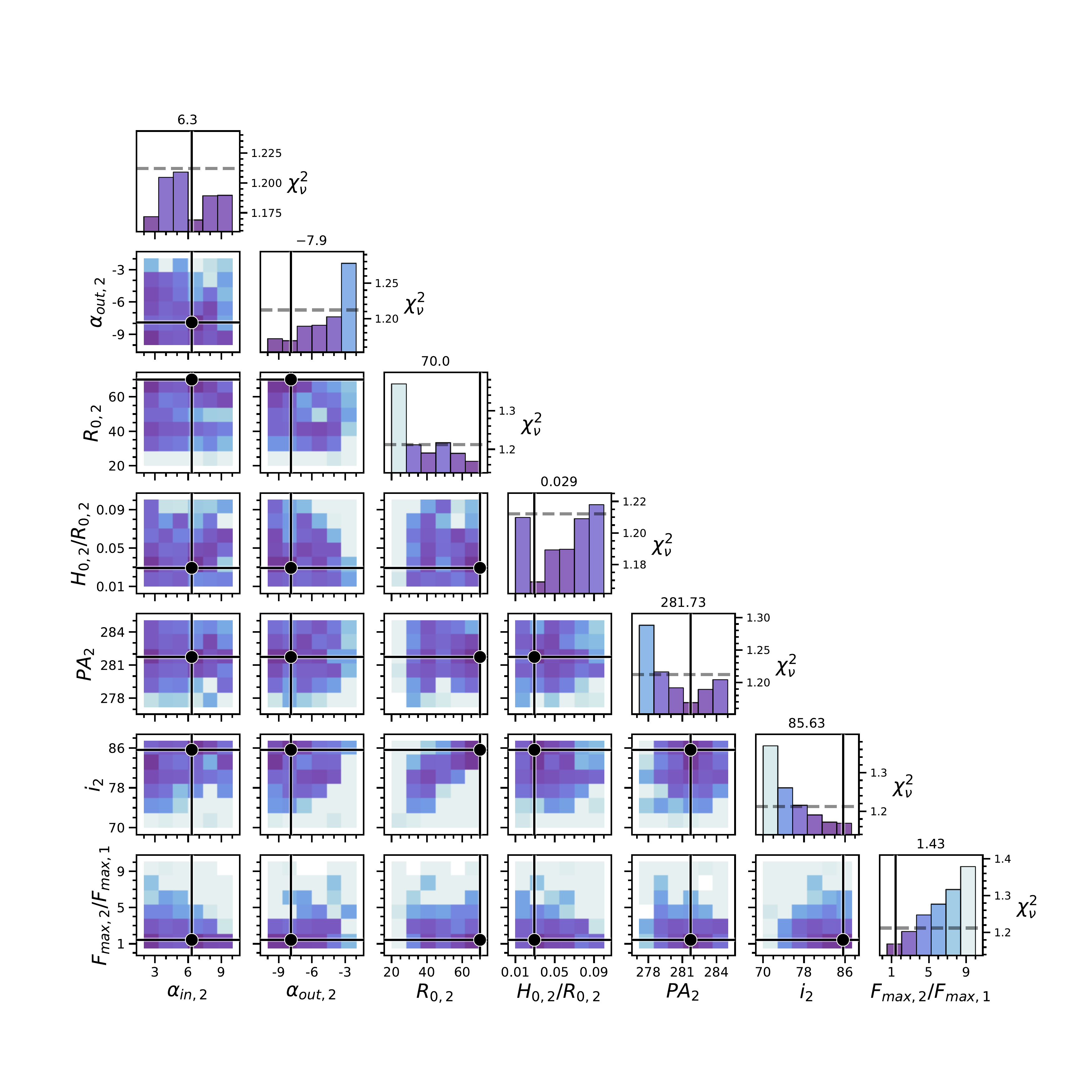}
\caption{As Figure \ref{fig:onecomp_corner} but for optimization of the two ring model.
\label{fig:twocomp_corner}
}
\end{figure*}

\textbf{Two Ring Models} -- 
Following the initial DE procedure, the optimal fit for two ring models results in $\chi^2_\nu = 1.169$ for the two A-LOCI reductions. A visualization of the full sample of models explored using differential evolution is provided in Figure \ref{fig:twocomp_corner}. The best solution at this stage has an inner ring radius that falls at the upper boundary of allowed values. We note, however, that the selected bounds include the best-fit inner rings of both \citet{MacGregor2019} ($\rm{R}_{0,2} \sim 48$ au) and \citet{Engler2019} ($\rm{R}_{0,2} = 64$ au). 

From the 910 models evaluated here, 20 resulted in acceptable values of $\chi^2_\nu$ ($\chi^2_\nu \leq 1.212$ with $\nu = 1076$). After KLIP forward modeling for this subset, the final best model resulted in $\chi^2_{\nu,\rm{tot}} = 1.151$ with a revised acceptable limit of $\chi^2_{\nu,\rm{tot}} \leq 1.181$ ($\nu = 2160$). Of the 20 models evaluated for all four reductions, 12 pass this revised threshold for acceptability. For clarity, we note that the best model from the initial DE/A-LOCI only reductions (the model whose parameters are indicated in Figure \ref{fig:twocomp_corner}) is ultimately excluded by this final acceptability threshold. This model simply ends up being a worse explanation for the KLIP results than other acceptable models, which pushes its combined score up sufficiently to be eliminated (with $\chi^2_{\nu,\rm{tot}} = 1.183$). The parameters for the best overall model and ranges of acceptable parameters are included in Table \ref{tab:twocomp_results}.

    \begin{deluxetable}{cccc}[htb!]
    \tablecaption{One Ring Model Optimization Results}
    \tablehead{\colhead{Parameter} & \colhead{Bounds}
    & \colhead{Best} & \colhead{Acceptable$^{a}$}}
    \startdata
    $R_{0}$ (au) & 86.0 -- 106.0 & 93.2 & 87.7 -- 99.6 \\
    $\alpha_{in}$ & 2.0 -- 10.0 & 2.0 & 2.0 -- 3.6 \\
    $\alpha_{out}$ & -10.0 -- -2.0 & -9.6 & -10.0 -- -7.5 \\
    $H_{0} / R_{0}$ & 0.01 -- 0.05 & 0.05 & 0.04 -- 0.05 \\
    $\rm{PA}$ (deg) & 277.0 -- 281.0 & 279.8 & 279.4 -- 280.5 \\
    $\rm{i}$ (deg) & 85.0 -- 87.0 & 85.3 & 85.3 -- 85.7 \\
    \tableline
    $\chi^2_{\nu,\rm{tot}}$ &   ---  &   1.166  & $\leq 1.196$ \\
    \enddata
    \tablecomments{Optimization bounds, best fitting value, and the range of acceptable values for each varied parameter of the single ring model following propagation through A-LOCI and KLIP forward modeling procedures for August 30 and September 07 data. $\chi^2_{\nu,\rm{tot}}$ indicates the combined measure for all four reductions (see Section \ref{sec:model_opt}). $a)$ These ranges give the smallest and largest value of each parameter that resulted in an acceptable solution. Given the possibility of complicated correlations between parameters and the lack of perfect sampling, it cannot be stated conclusively that every value within these ranges can produce an acceptable solution.\label{tab:onecomp_results}}
    \end{deluxetable}
    
    \begin{deluxetable}{cccc}[htb!]
    \tablecaption{Two Ring Model Optimization Results}
    \tablehead{\colhead{Parameter} & \colhead{Bounds}
    & \colhead{Best} & \colhead{Acceptable}}
    \startdata
    $R_{0,2}$ (au) & 20.0 -- 70.0 & 40.9 & 36.3 -- 62.1 \\
    $\alpha_{in,2}$ & 2.0 -- 10.0 & 3.0 & 2.0 -- 8.2 \\
    $\alpha_{out,2}$ & -10.0 -- -2.0 & -5.3 & -9.4 -- -4.6 \\
    $H_{0,2} / R_{0,2}$ & 0.01 -- 0.10 & 0.03 & 0.02 -- 0.08 \\
    $\rm{PA}_2$ (deg) & 277.2 -- 285.0 & 281.6 & 280.2 -- 283.8 \\
    $\rm{i}_2$ (deg) & 70.0 -- 87.3 & 80.1 & 79.0 -- 85.0 \\
    $F_{max,2} / F_{max,1}$ & 0.5 -- 10.0 & 2.0 & 1.2 -- 2.7 \\
    \tableline
    $\chi^2_{\nu,\rm{tot}}$ &  --- & 1.151 & $\leq 1.181$ \\
    \enddata
    \tablecomments{As Table \ref{tab:onecomp_results}, but for two ring model optimization.\label{tab:twocomp_results}}
    \end{deluxetable}

\begin{figure*}
\includegraphics[width=\textwidth]{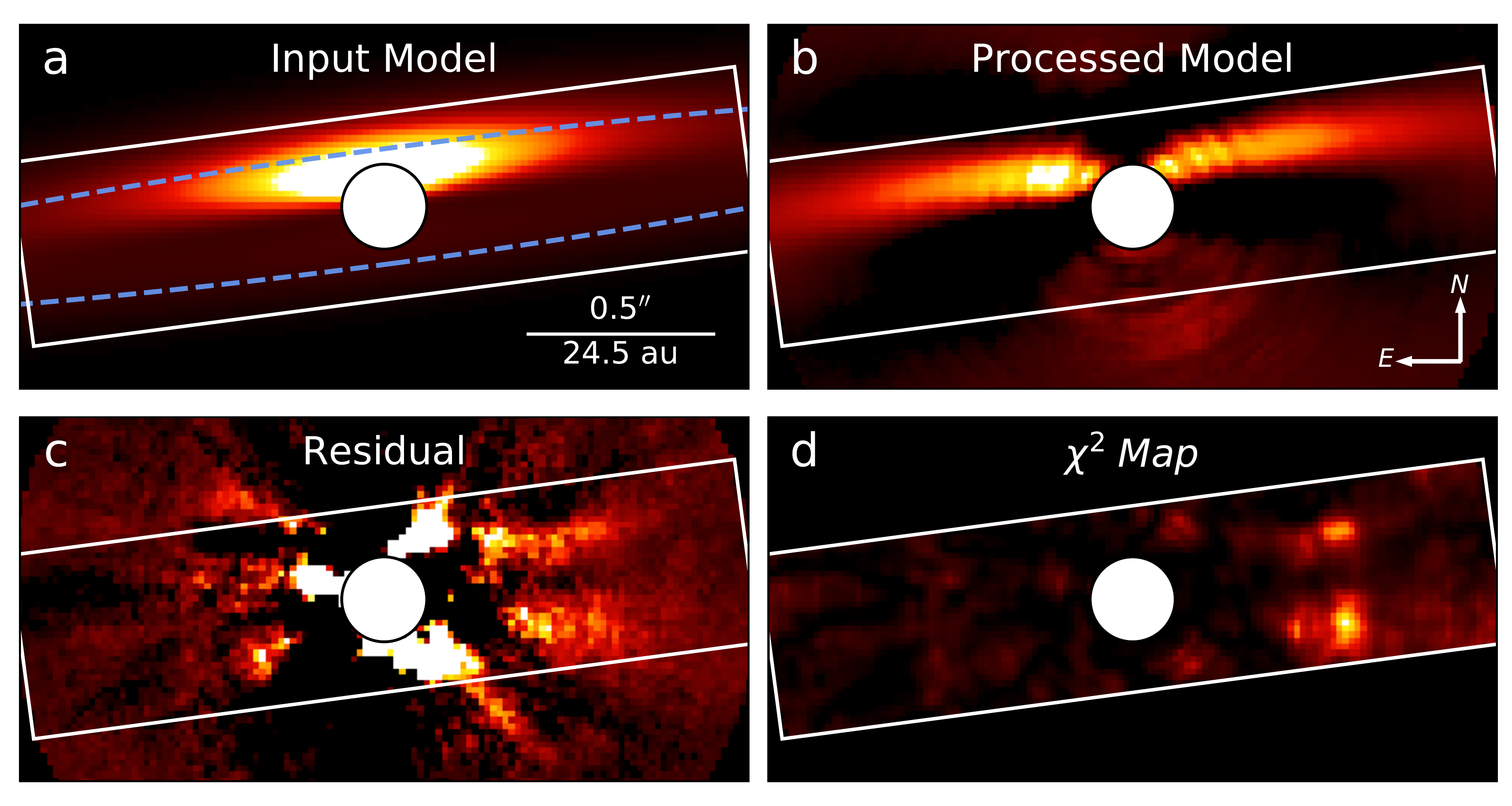}
\caption{
The lowest $\chi^2_{\nu,\rm{tot}}$ single ring model ($\chi^2_{\nu,\rm{tot}} = 1.166$) identified by the procedure outlined in Section \ref{sec:forward_modeling}, depicted for our KLIP reduction of August 30 data. Model parameters utilized can be found in Table \ref{tab:onecomp_results}. In all panels, the white rectangle indicates the ``region of interest'' for the purpose of $\chi^2$ calculation, with the white circle indicating the inner exclusion radius. \textbf{a}) the input disk model convolved with the instrumental PSF. A schematic of the disk model is overlaid as a dashed blue ellipse having the same radius, inclination, and PA as the model disk. \textbf{b}) the disk model after application of the forward modeling procedure for the August 30 2017 KLIP reduction, displayed exactly as the images of Figure \ref{fig:reducs} (linear stretch over $-0.61-6.10$ $mJy$ $arcsec^{-2}$). \textbf{c}) Residuals for the August 30 2017 KLIP data product (upper left image in Figure \ref{fig:reducs}) and the processed disk model, also displayed as Figure \ref{fig:reducs} (linear stretch over $-0.61-6.10$ $mJy$ $arcsec^{-2}$). \textbf{d}) unbinned $\chi^2$ map for this model displayed with linear stretch over 0 to 19.46 in $\chi^2$ (roughly equivalent to $0-0.036$ in $\chi^2_\nu$).
\label{fig:onecomp_model}
}
\end{figure*}

\begin{figure*}
\includegraphics[width=\textwidth]{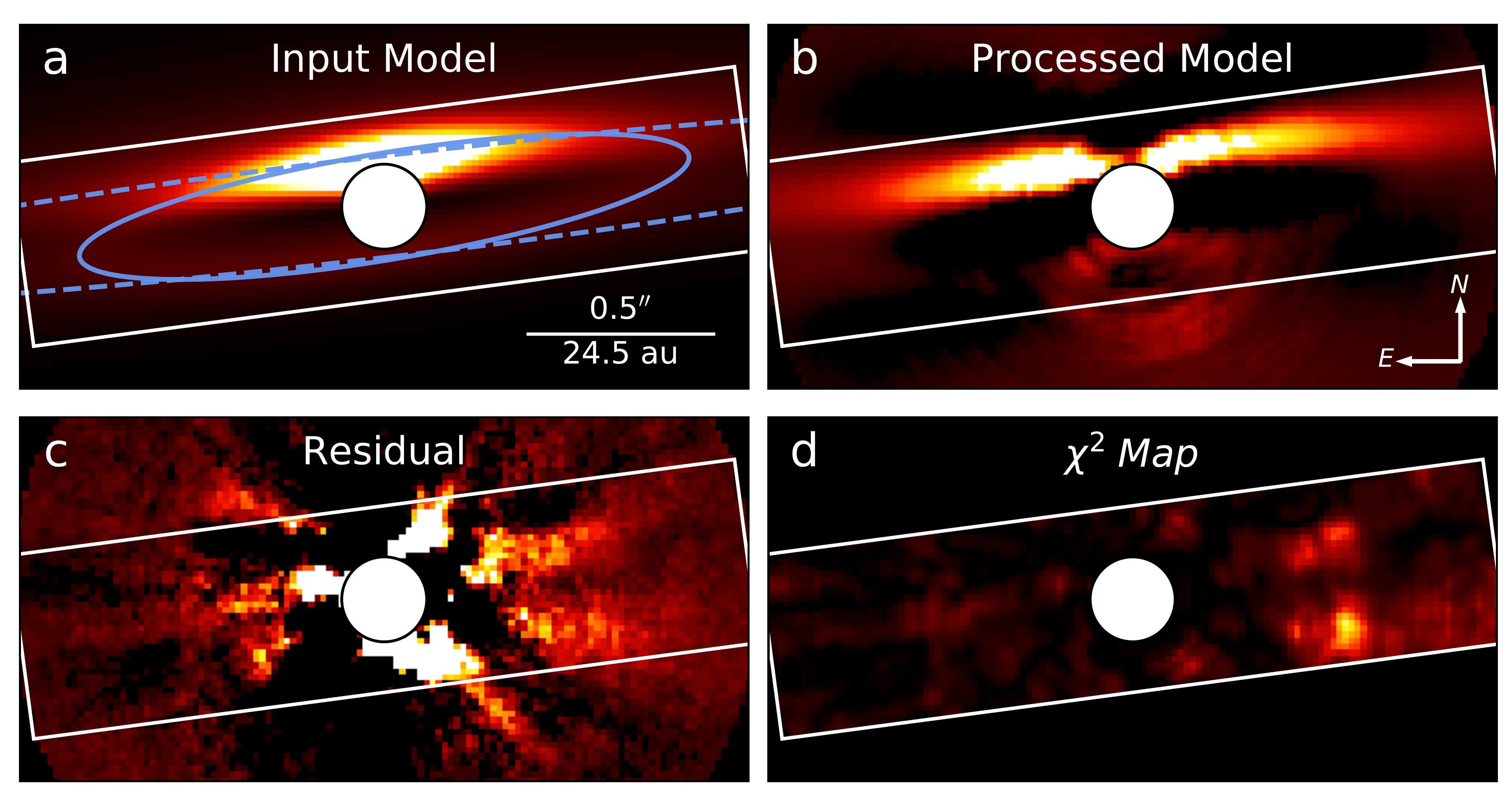}
\caption{
The lowest $\chi^2_{\nu,\rm{tot}}$ two ring model ($\chi^2_{\nu,\rm{tot}} = 1.151$) identified by the procedure outlined in Section \ref{sec:forward_modeling}. The schematic overlaid in panel $a)$ shows the inner ring as a solid blue ellipse, and the outer as a dashed blue ellipse. The $\chi^2$ map shown in panel d) is displayed with identical scaling to that of Figure \ref{fig:onecomp_model}. Otherwise, as Figure \ref{fig:onecomp_model}.
\label{fig:twocomp_model}
}
\end{figure*}

\subsection{Modeling Discussion}\label{sec:disc_modeling}
With the optimal one and two ring models identified producing comparable $\chi^2_\nu$ metrics of 1.166 and 1.151 respectively, both geometries appear statistically consistent with our data -- making it difficult to rule out either scenario. However, our exploration of the model parameter spaces using differential evolution allows us to place some disambiguating constraints.

A few noteworthy observations can be made regarding the one ring model results. Firstly, the corner plot of our differential evolution procedure (Figure \ref{fig:onecomp_corner}) reveals a preference for a very slow density change interior to $R_0$ (small $\alpha_{in}$), with a very rapid change exterior to $R_0$ (large $\alpha_{out}$). For both of these parameters, the optimization converges at, or very near to, the boundaries (2 and -10 respectively). While this could suggest true values for these parameters beyond the boundaries we've set, this seems unlikely given the results of previous studies of HD 15115's disk; for example, the observed extent of the disk with wider fields of view is inconsistent with a radial density profile having $\alpha_{out} \sim -10$. We note, however, that models more consistent with prior results produce acceptable fit metrics as well (see Table \ref{tab:onecomp_results}). Further, given our narrow field-of-view and the disk's high inclination, we should expect that our results are not as sensitive to changes in these particular parameters anyway. Additionally, the PA identified is seemingly distinct from values typically found in previous studies of the disk, with our procedure finding $\rm{PA} = 279\fdg8 ^{+0.7}_{-0.4}$ \footnote{Uncertainties here are roughly approximated as simply the upper and lower limits for acceptable models as presented in Table \ref{tab:onecomp_results}.} compared to $\rm{PA} = 278\fdg9 \pm 0\fdg1$ from \citet{Engler2019}. However, the value identified here is consistent with the value of $280.02\pm0.27$ identified from ellipse fitting of the spine in Section \ref{sec:disk_spine}.

\begin{figure}
\includegraphics[width=8cm]{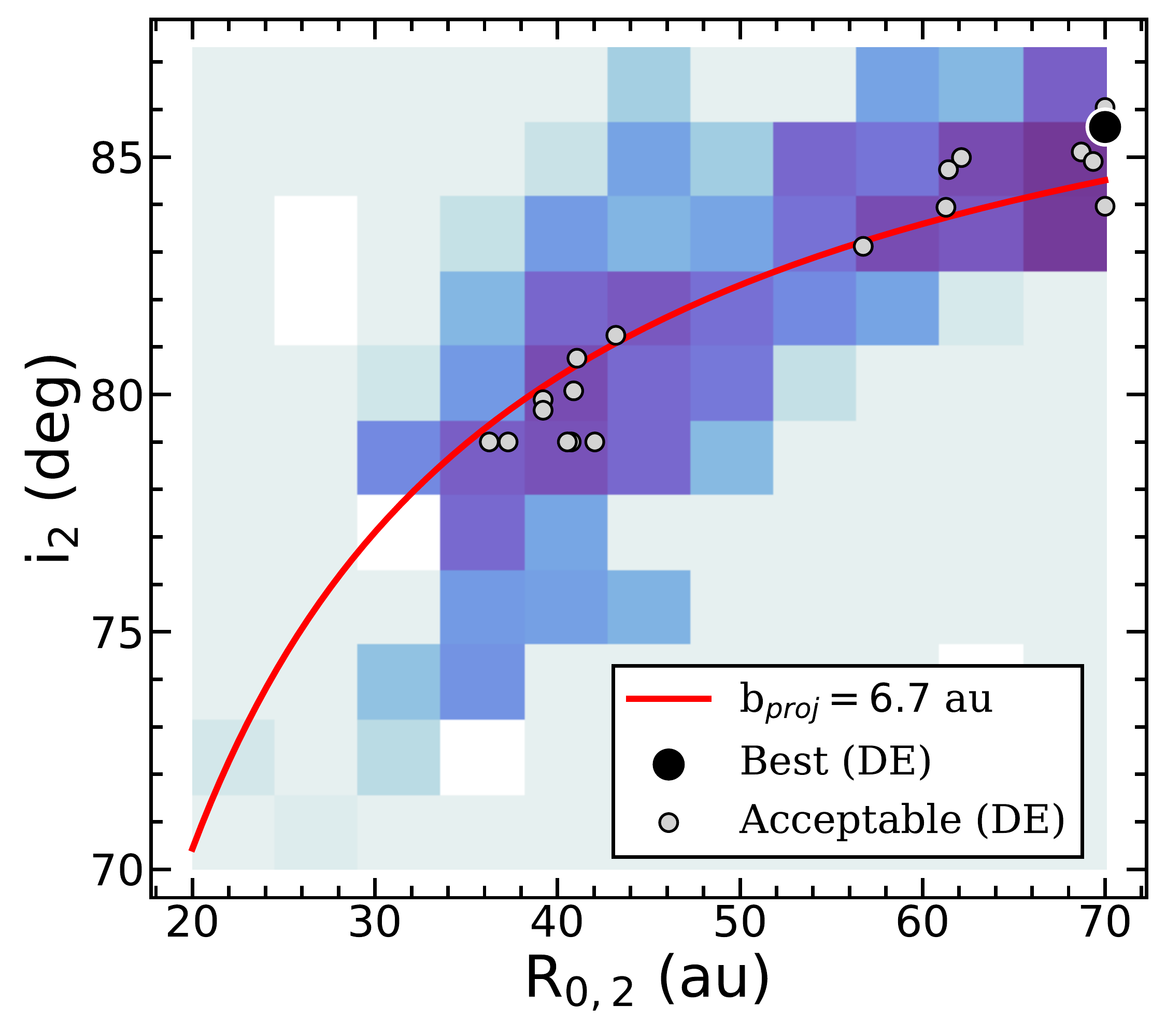}
\caption{Reproduced from Figure \ref{fig:twocomp_corner}: best $\chi^2$ for two ring models as a function of inner disk radius and inclination. Values corresponding to the best model and acceptable models following the DE procedure of Section \ref{sec:forward_modeling} (i.e. $\chi^2_\nu$ accounting for only the A-LOCI reductions) are shown in black and grey respectively. The red line indicates values of radius and inclination that produce a projected semi-minor axis equal to that of the adopted outer ring. Acceptable two ring models are identified only for inner disk parameters resulting in nearly the same projected semi-minor axis (see Section \ref{sec:disc_modeling}).
\label{fig:semi_minor_axis}
}
\end{figure}

From the schematic of our best two ring model (Figure \ref{fig:model_schematics}, top panel), we see that our data is best explained by an inner ring with a projected semi-minor axis similar to that of the outer ring ($b_{proj} \sim 6.7 \; au$ for $R_0 = 96 \; au$ and $\rm{i} = 86\degr$). Looking at the corner plot for the two ring optimization (Figure \ref{fig:twocomp_corner}), perhaps the strongest apparent correlation occurs between the inner ring's radius (equivalently its projected semi-major axis, a) and its inclination. Noting that the projected semi-minor axis, b, of the inner ring is related to its inclination and projected semi-major axis by $b_{proj} = a_{proj}\cdot cos(\rm{i})$, the correlation in $R_{0,2}$ versus $\rm{i}_2$ subplot falls very nearly along the line corresponding to $b_{2,proj} = 6.7 \;au$ (see Figure \ref{fig:semi_minor_axis}). In fact, for the full set of acceptable two ring models following differential evolution, all have projected semi-minor axes between 4.8 and 8.0 au -- while values from 0.9 to 23.9 au are permitted by our parameter bounds. The revised set of acceptable models following modeling of all four CHARIS reductions reduces this range even further, to 5.4 -- 8.0 au. Combined with our analysis of the disk's spine and surface brightness profiles, our results appear to suggest a lack of any statistically significant distinct inner ring spine.

Overall, the results of our modeling procedure can be interpreted in a number of ways:

\begin{enumerate}
\item From the strong preference for $b_{1,proj} \sim b_{2,proj}$: a distinct inner ring exists but its brightest features happen to roughly line up along our line of sight with the canonical outer ring. This would result in a two ring geometry that is statistically indistinguishable from one with a single ring. A distinct inner ring that is coplanar (or nearly coplanar) with the outer ring is consistent only as the inner ring's radius approaches that of the outer, where $b_{2,proj}$ can near $b_{1,proj}$ while maintaining a matching inclination.
\item In the CHARIS field of view ($\sim 0\farcs2 - 1\farcs0$) either the outer or inner ring is substantially brighter, such that the other is not recovered in our data. The differing PA identified for the one ring model compared to literature could suggest that we're seeing the latter. However, we remark that our optimization procedure for two ring models allowed values of $F_{max,2} / F_{max,1}$ as large as 10; if the underlying system is well-described within CHARIS's field of view as an extremely faint outer ring with a misaligned bright inner ring, solutions with large ring flux ratios should have manifested.
\item Perhaps the system is truly better described as a single ring. The tendency of the single ring DE procedure toward parameter values which are seemingly at odds with prior results could be caused by inaccuracies in assumptions made by our models. e.g. if the true scattering phase function departs slightly from the Hong phase function assumed, if the disk is non-negligibly eccentric, or if the disk features a non-linear flaring profile, models matching the true underlying parameters may not coincide with the minimum $\chi^2_\nu$ in our analysis. In this case, even if no second ring exists, a second ring model might serve to mitigate these inaccuracies sufficiently to result in competitive fitness metrics.
\item The disk has a geometry distinct from any probed here and that is difficult to diagnose as a result of the nearly edge-on orientation (e.g. a debris disk with spiral arms or significant warping).
\end{enumerate}

We also point out the presence of the very small separation ($\rho \sim 0\farcs25$) residual signal that is not fit by our models (bright residuals appearing just beyond the software mask in panel c of Figures \ref{fig:onecomp_model} and \ref{fig:twocomp_model}). While this feature could be evidence of the $\sim 4$ au ``warm-dust" disk suggested by \citet{Moor2011} through SED fitting, perhaps the more likely explanation is that it is simply residual speckle noise. While ``aggressive processing" is often cited as a cause of spurious features in debris disks \citep{Duchene2020}, poor speckle suppression due to \textit{insufficiently} aggressive PSF subtraction can also cause spurious features at small angular separations. Indeed, separate tests with more aggressive A-LOCI and KLIP settings for our data appear to confirm that this signal is due to residual speckles, but these approaches compromise our detections of the disk at 0\farcs{}25--1\arcsec{} and thus are not used. 

\subsection{Comparison with Recent Studies}

Based on SPHERE polarimetry, \citet{Engler2019} suggest the possibility of a misaligned/non-coplanar inner ring with a radius of $\sim 1\farcs3$ -- ultimately finding a slightly better fit to their non-polarized data for this geometry than for a one-ring or coplanar two-ring geometry. They further investigate the merit of the additional parameters of the two-ring model by comparing the Bayesian Information Criterion (BIC) for the two models, concluding that the BIC metric supports their best-fit non-coplanar two-ring model. While our modeling shows that CHARIS data is consistent with a misaligned inner ring for some combinations of inclination and PA (see Sections \ref{sec:disc_modeling} and \ref{sec:model_results}), we find no clear evidence indicating the presence of an inner ring oriented as hypothesized from SPHERE imagery ($\rm{PA} = 276\degr$, $\rm{i}=80\degr$; see bottom panel of Figure \ref{fig:model_schematics}). Carrying out forward modeling on the best-fitting misaligned two ring model identified in \citet{Engler2019} appears to reinforce this, with the model producing a $\chi^2_{\nu, \rm{tot}}$ of 2.13 (for the same assumption of M=10 free model parameters that they indicate) after forward modeling for our four reductions, compared with 1.151 for the overall best two ring model we identify (see fourth row of Figure \ref{fig:phase_funcs}). Even allowing freedom for the other parameters, our two ring optimization identified no strong solutions having the inner ring oriented similarly (see Figure \ref{fig:twocomp_corner}). The best coplanar two ring model they identify fits our data somewhat better, resulting in a $\chi^2_{\nu, \rm{tot}}$ of 1.7 (see bottom row of Figure \ref{fig:phase_funcs}). Models with a similar inner ring radius and roughly coplanar orientation manifest in our final set of acceptable models (Table \ref{tab:twocomp_results}) when using the Hong scattering phase function instead. 

\begin{figure*}
\includegraphics[width=\textwidth]{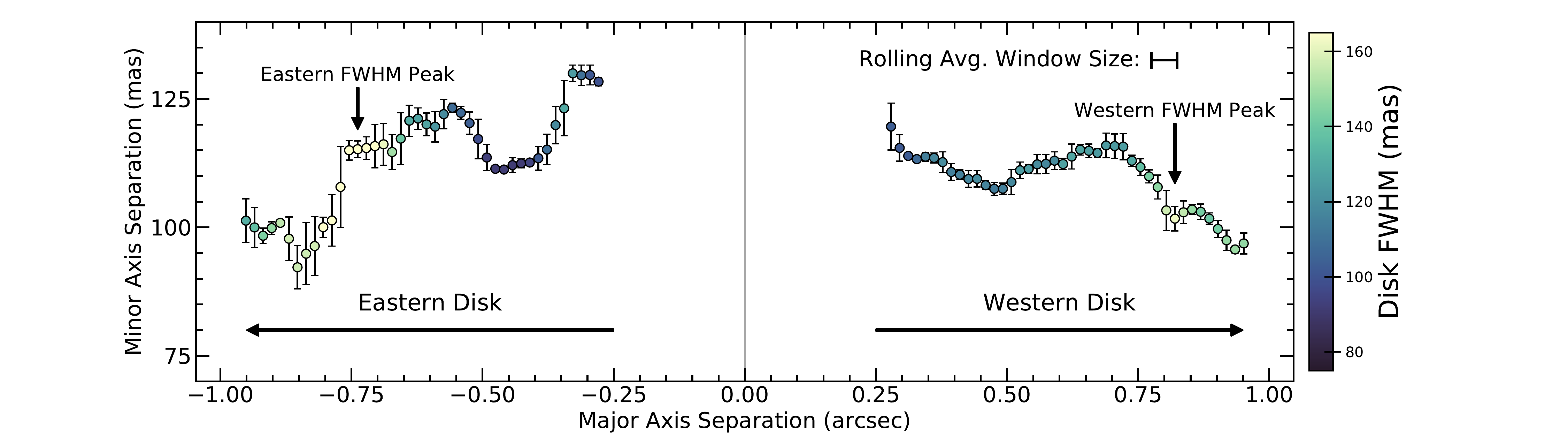}
\caption{Disk spine position and projected FWHM measurements made in Section \ref{sec:disk_spine} with a 3 pixel wide rolling weighted average applied. At $\sim 0\farcs75 - 0\farcs80$ on either side, a $\sim 1-2\sigma$ dip in minor axis separation is coincident with an increase in measured disk FWHM. Further, a peak in FWHM can be seen on both sides, but occurs slightly asymmetrically. 
\label{fig:smoothed_spine}
}
\end{figure*}

\begin{figure*}
\includegraphics[width=\textwidth]{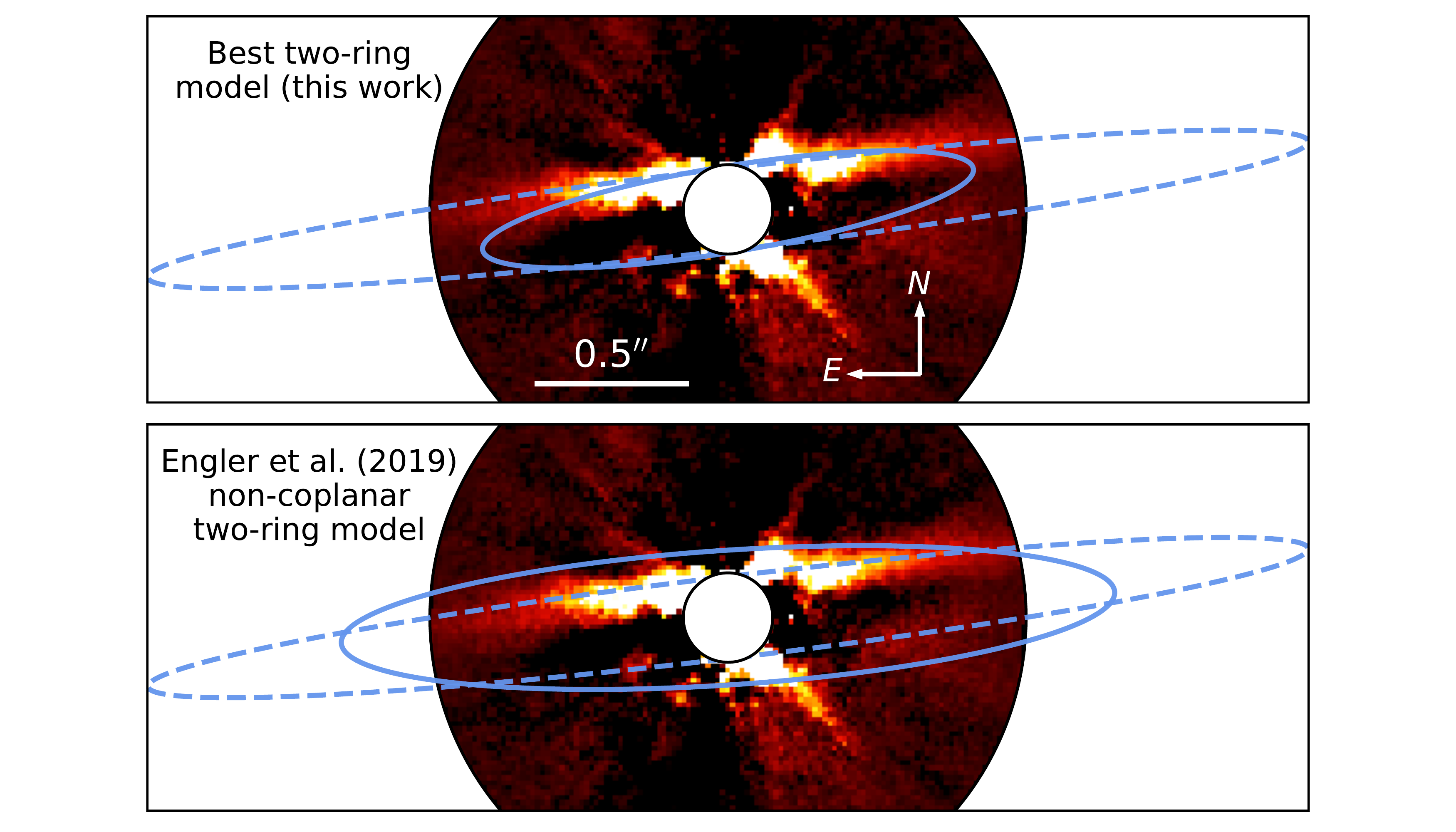}
\caption{\textbf{Top:} Image of A-LOCI processed Aug 30 CHARIS data with overlaid ellipses corresponding to the outer (dashed) and inner (solid) rings of our best-fit two-ring solution (see Section \ref{sec:models}). While our solution's inner ring has an inclination offset of 5\fdg9 and a PA offset of 2\fdg7 with respect to the outer ring, the best-fit inner ring's radius results in an inner ring whose features predominantly coincide with those of the outer ring along our line of sight. \textbf{Bottom:} As above, but with schematic depicting the inner and outer rings of the best-fit misaligned two-ring geometry from the results of \citet{Engler2019} (see also: the second row from the bottom of Figure \ref{fig:phase_funcs}). An inner ring oriented as posited by \citet{Engler2019} is not evident in the CHARIS data; to exist at such an orientation and still be consistent with our data, it would need to be substantially dimmer than the outer ring at similar projected separations.
\label{fig:model_schematics}
}
\end{figure*}

From ALMA observations, \citet{MacGregor2019} favor a disk model composed of two coplanar rings or a single ring with a Gaussian gap. Given the differences in parametizations between our models, it is difficult to unambiguously translate their results for direct application to our data. However, the gap suggested by their models in either case is small enough ($\sim 14$ au) that the profile should manifest consistently with the profile we observe (e.g. with the appearance of a single spine in our imagery; see Section \ref{sec:disc_modeling} for relevant discussion regarding this constraint).  We also note that our spine trace (Figure \ref{fig:spine}) shows a $\sim 1\sigma$ shift around 0\farcs8 on either side which is roughly coincident with the inner edge of the inner ring they propose. By applying a simple 3-pixel rolling weighted average to the spine trace and disk projected FWHM measurements, this feature becomes more clear (see Figure \ref{fig:smoothed_spine}). While we find no significant evidence to support the presence of the $\sim 14$ au gap that they favor (given that it falls outside of our field of view, with $\rho \sim 1\farcs2$), their interpretation appears generally consistent with CHARIS imagery.

The smoothed disk spine in Figure \ref{fig:smoothed_spine} also appears remarkably similar to the single profile spine trace of NZ Lup's disk (another highly inclined debris disk) reported in \citet{Boccaletti2019} (their Figure 3). \citet{Boccaletti2019} ultimately favor a mutually inclined ($\Delta i \sim 5\degr$) two ring model for NZ Lup with an apparent gap that is roughly coincident with the dip in spine position seen in their spine trace. If a comparable explanation is assumed for the $\sim 0\farcs75 - 0\farcs80$ feature in our smoothed spine trace, the result would be a two ring disk with a gap similar to the one suggested by \citet{MacGregor2019}, but at a somewhat smaller separation than their best fit. Though, given that the ALMA observations trace a significantly different dust population than ours, these results may be fully consistent with one another. Notably, this interpretation manifests similarly to the overall best two ring model identified in Section \ref{sec:model_results} (see Table \ref{tab:twocomp_results}), with a fiducial inner ring radius of 40.9 au ($0\farcs83$) and a mutual inclination of 5\fdg9. The slight difference in the location of peak FWHM measured for the disk between the east and west sides in Figure \ref{fig:smoothed_spine} ($\sim 0\farcs75$ and $\sim 0\farcs82$ respectively) might be explained by the difference in PA between the rings suggested by our best two ring model ($\Delta \rm{PA} = 2\fdg7$).

\section{Disk Surface Brightness and Color}\label{sec:color}
\subsection{Disk Photometry}
\begin{figure*}
\includegraphics[width=\textwidth]{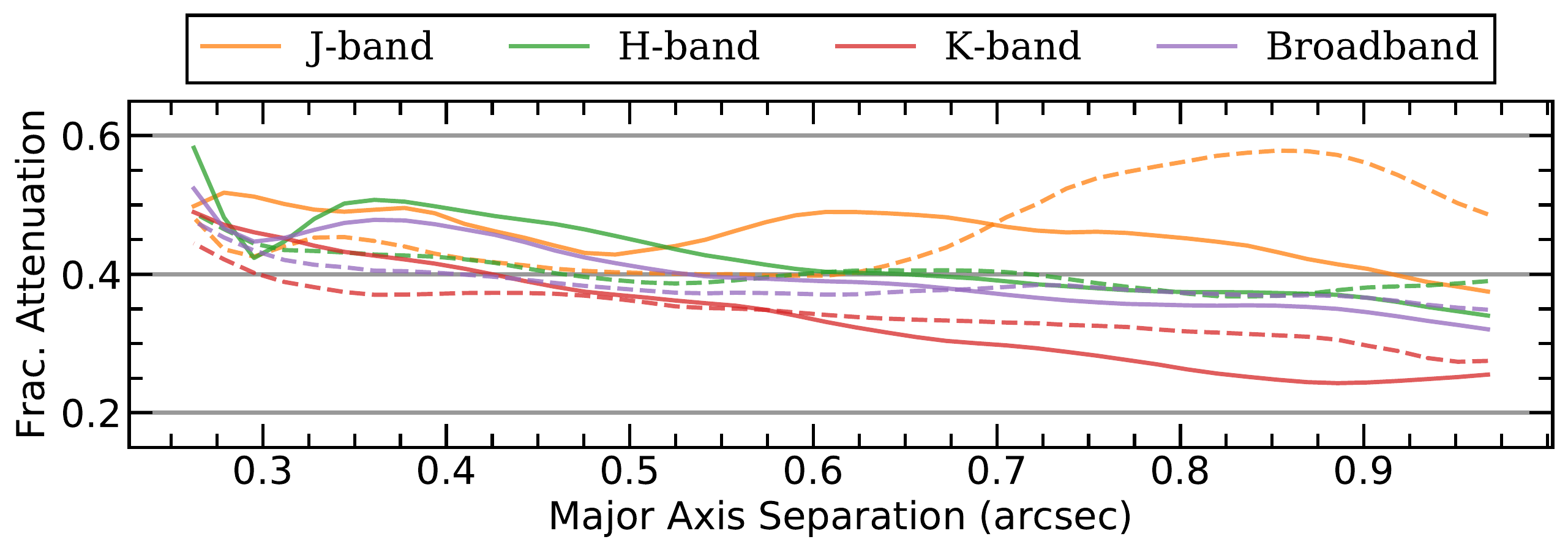}
\caption{Estimates of fractional disk flux attenuation as a function of separation along the major axis. Solid and dashed lines show the values along the western and eastern extents of the disk respectively. Values manifest similarly in all four reductions, but are shown here for our A-LOCI reduction of August 30 data. These values correspond to $1-\frac{1}{C}$, where C is the effective attenuation correction applied during measurement of surface brightness (see Section \ref{sec:color})}.
\label{fig:atten_fact}
\end{figure*}

To analyze the brightness and color of HD 15115's disk, we follow the general procedure of \citet{Goebel2018} to produce surface brightness profiles in CHARIS broadband and J, H, and K bands. However, instead of fitting a fourth order polynomial to the identified disk spine (see Section \ref{sec:disk_spine}), we adopt the positions of the best-fit ellipse as the location of the disk spine for all imagery utilized \footnote{Though different observing wavelengths may trace distinct dust populations, resulting in different spine positions, testing showed that the utilized positions fall very near to locations we fit for HST/STIS imagery where meaningful fitting was feasible. For the purpose of surface brightness measurements, the spine identified from CHARIS broadband imagery appears to be a reasonable approximation of the spine for all bands we analyze.}. For measuring surface brightness, imagery is not rotated as it was in Section \ref{sec:disk_spine}. Rather, the spine locations measured for the rotated images are transformed to the native (north-up) image orientations, eliminating the possibility of image rotation interpolation affecting surface brightness measures. Flux attenuation cubes are created by dividing the PSF subtracted best disk model by the pre-PSF subtracted disk model (see: Section \ref{sec:forward_modeling}). \citet{Bhowmik2019} show that a comparable procedure results in erratic attenuation estimates for their KLIP reduction of SPHERE data for the highly inclined debris disk system HD 32297. However, this behavior does not manifest in our case (see Figure \ref{fig:atten_fact}). These attenuation measurements are then used to correct the PSF subtracted cubes produced with A-LOCI and KLIP reductions of August 30 and September 07 data. As in \citet{Goebel2018}, we see fractional attenuation that tends to increase at smaller separations and further from the disk spine. Along the spine, disk flux in CHARIS broadband is attenuated by $\sim 45-55\%$ at 0\farcs25 separation, and by $\sim 30-35\%$ at 0\farcs75. This attenuation varies by wavelength, with J-band typically being most attenuated ($\sim 65 \%$ and $45\%$ at 0\farcs25 and 0\farcs75 resp.), followed by H-band (with values comparable to those in broadband), and with K-band suffering the least attenuation ($\sim 45 \%$ and $30\%$ at 0\farcs25 and 0\farcs75 resp.). 

Following this, the channels of the attenuation-corrected reduction products are merged to create images corresponding to J (channels $1-5$, $1.16-1.33$ $\micron$), H (channels $8-14$, $1.47-1.80$ $\micron$) and K (channels $16-21$, $1.93-2.29$ $\micron$) bands. For each (x,y) pixel position along the spine, we take the nominal surface brightness to be the average flux value within a circular aperture with diameter 0\farcs12 (approximately the narrowest observed disk FWHM in CHARIS broadband imagery; this aperture size is used for all imagery). To accommodate the inclusion of partial pixels, we take the average in an aperture to be the weighted mean of the values with weights equal to the exact fraction of each pixel that is included in the aperture. 

The uncertainty for each surface brightness measurement is determined as follows using non-attenuation-corrected images (attenuation maps become extremely noisy at the faint edges of the disk, where both attenuated and unattenuated models have values near zero). If the surface brightness, $F_{s}$, is measured at spine position $(x_s, y_s)$ with corresponding stellocentric polar coordinates $(r_s, \theta_s)$, we make additional measurements the same way within apertures at an array of positions $(r_s, \theta_i)$, with $\theta_i$ placed every $10\degr$. Any of these measurements whose aperture overlaps with any spine aperture are removed. We then compute the standard deviation of this array of measurements as $\sigma_{0,s}$. Since $\sigma_{0,s}$ is representative of the uncertainty in the average surface brightness for the aperture at position $(x_s, y_s)$ \textbf{before} applying the attenuation correction, we additionally compute the nominal surface brightness at $(x_s, y_s)$ in the uncorrected image, $F_{0,s}$. The effective attenuation correction applied is then $C_s = F_{s} \; / \; F_{0,s}$. From this, our final uncertainty for $F_s$ is taken to be $\sigma_s = C_s \cdot \sigma_{0,s}$.

\begin{figure*}
\includegraphics[width=\textwidth]{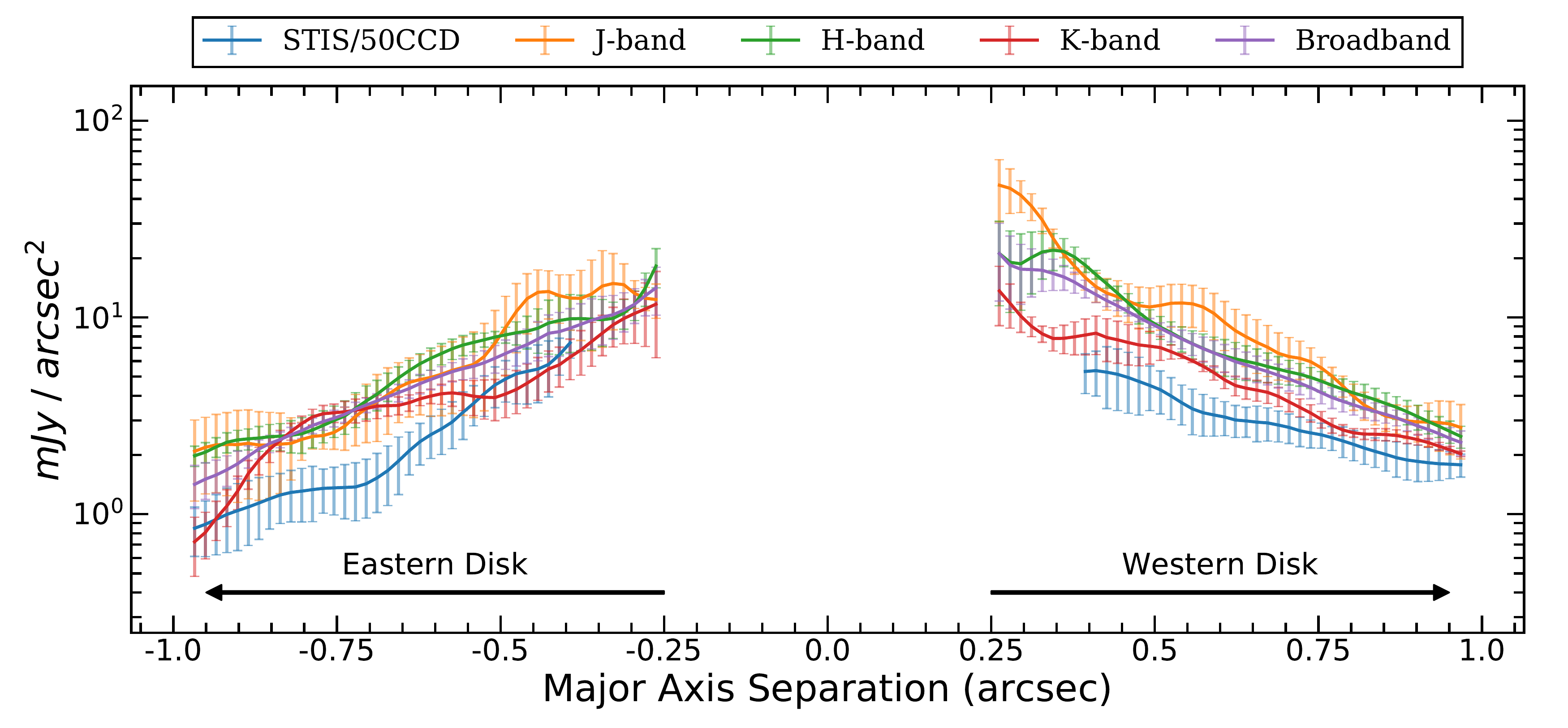}
\caption{Measurements of surface brightness for HD 15115's disk as a function of stellocentric separation along the major axis in CHARIS and HST/STIS data. Here (and for other figures from Section \ref{sec:color}), ``broadband'' refers to CHARIS's broadband. We note that the measurements here and in other Section \ref{sec:color}) figures are not contiguously independent as a result of the choice to use a single aperture size for all photometric bands analyzed.
\label{fig:sb}
}
\end{figure*}

The procedure above is repeated to get arrays of surface brightness and corresponding uncertainties for each filter and reduction. For a given filter, the final nominal surface brightness at each position is taken to be the inverse variance weighted average of the corresponding surface brightness measurements for each of the four reductions utilized. The corresponding uncertainty is taken to be the standard error on the weighted average (see footnote, Section \ref{sec:disk_spine}). Surface brightness is measured in the same manner for HST/STIS imagery of the system, except that no attenuation correction is necessary and only one reduction is used. The results of this procedure are depicted in Figure \ref{fig:sb}. We point out here that, although the disk appears to be recovered only marginally in J-band imagery (Figure \ref{fig:jhkbands}), making SB measurements over a large (0\farcs12) aperture and averaging measurements for multiple reductions results in J-band surface brightness measurements with reasonable signal-to-noise.

\begin{figure*}
\includegraphics[width=\textwidth]{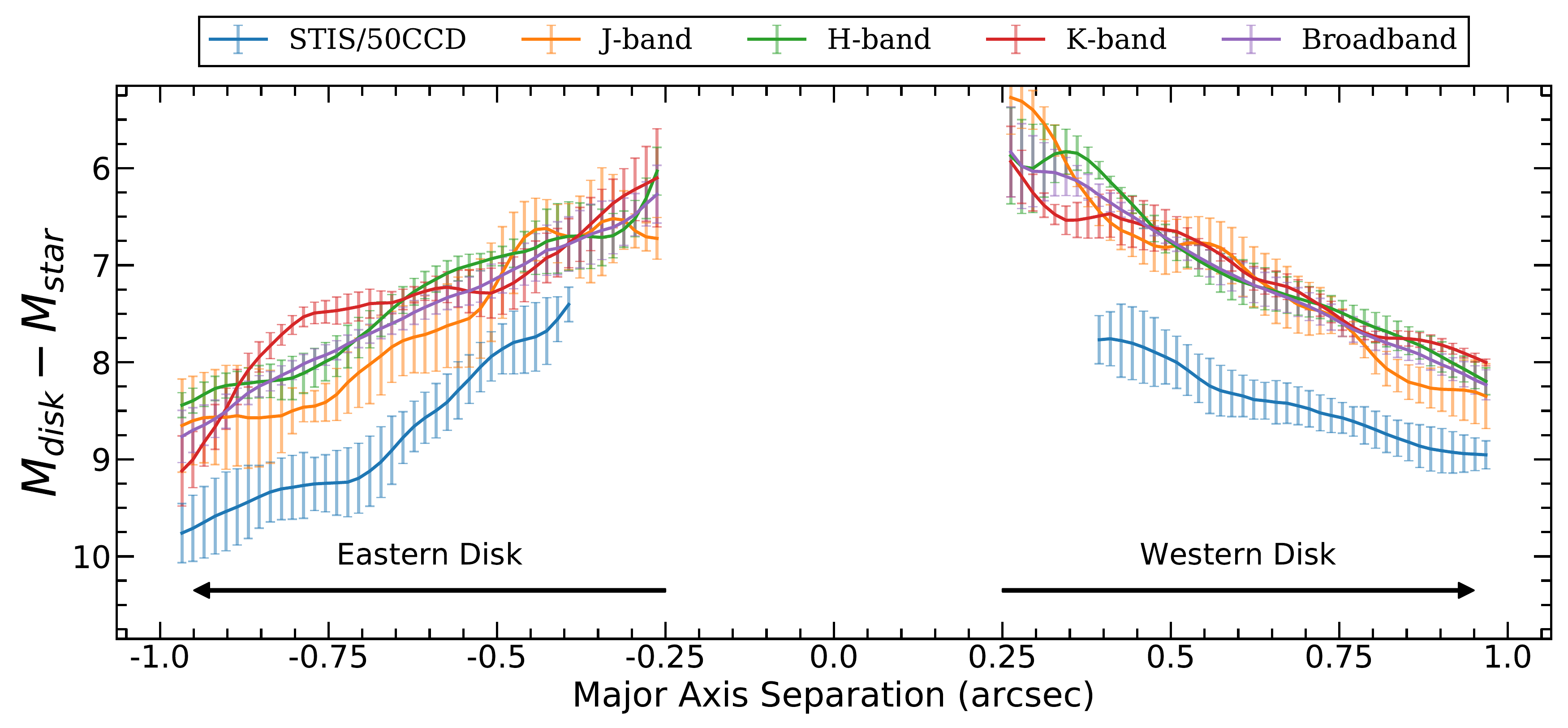}
\caption{Measurements of surface reflectance for HD 15115's disk in CHARIS and HST/STIS data.
\label{fig:sb_dmags}
}
\end{figure*}

The surface brightness measurements for each bandpass are then combined with measurements of the stellar flux (for CHARIS data: from analysis of satellite speckles during spectrophotometric calibration of the cubes, and for HST/STIS data: as reported in \citep{Schneider2014}) to compute the local surface brightness of the disk relative to the stellar flux (see Figure \ref{fig:sb_dmags}).

\begin{figure*}
\includegraphics[width=\textwidth]{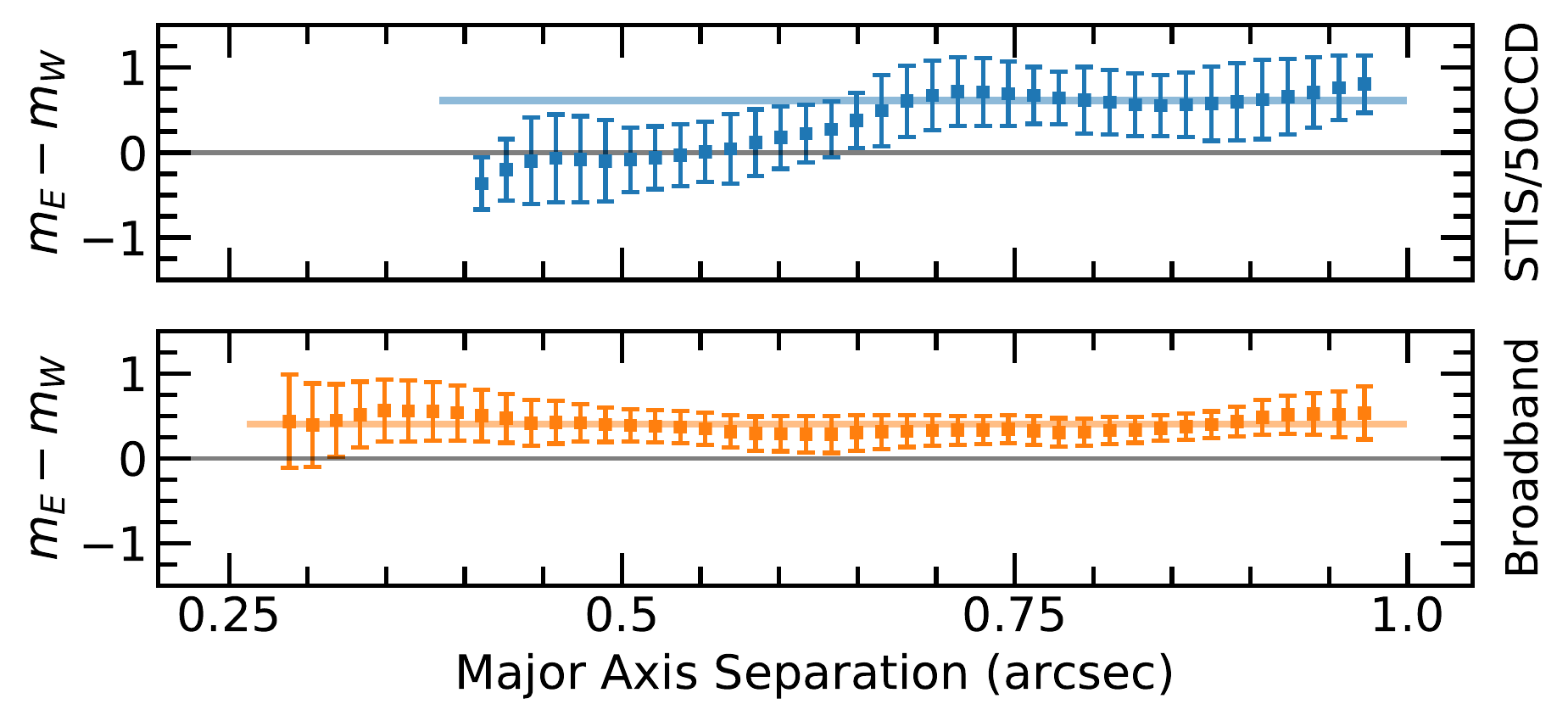}
\caption{Relative surface brightness between the eastern and western extents of the disk for STIS/50CCD and CHARIS broadband. In each subplot, the horizontal colored line indicates the weighted average of the constituent flux measurements. The asymmetry in J, H, and K manifests similarly to the CHARIS broadband, albeit with larger uncertainties.
\label{fig:sb_asymmetry}
}
\end{figure*}

\begin{figure*}
\includegraphics[width=\textwidth]{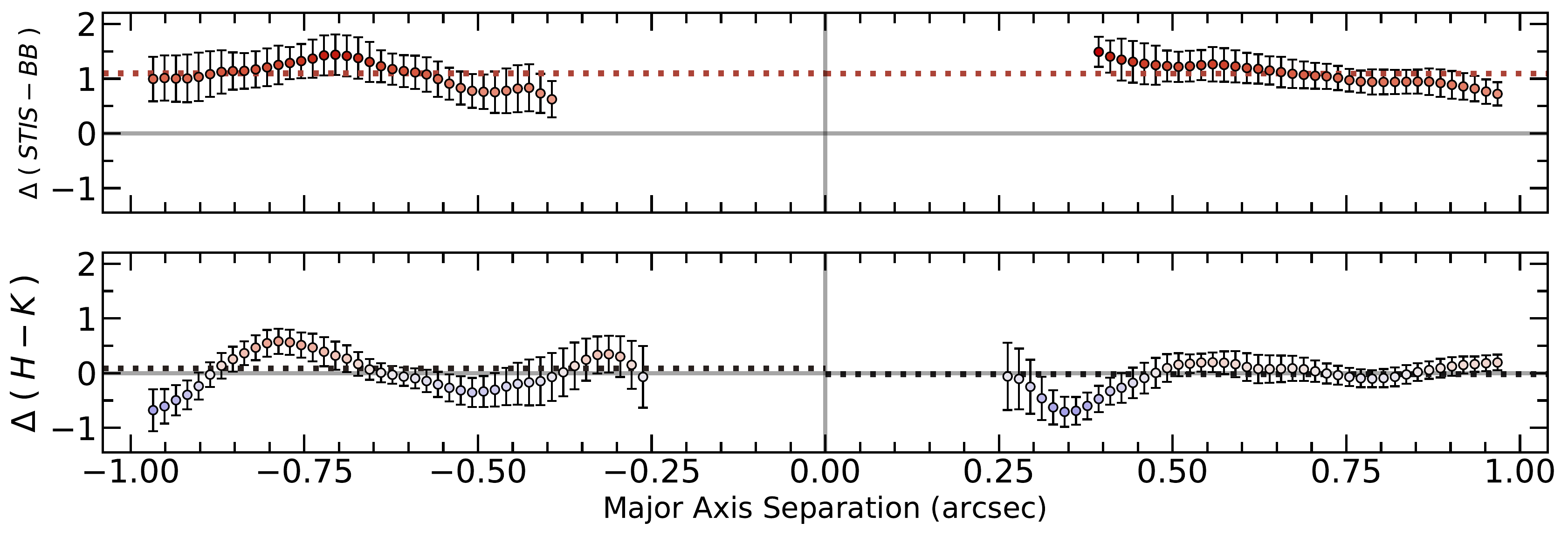}
\caption{Disk color as a function of stellocentric separation along the disk major axis as identified from Section \ref{sec:disk_spine} (see Section \ref{sec:color}), with $\Delta(H-K)=(H-K)_{disk}-(H-K)_{star}$, etc. Here, ``STIS'' refers to meaurements in STIS/50CCD, while ``BB'' refers to measurements in CHARIS's broadband (wavelength-collapsed) imagery. The horizontal dotted line in each subplot indicates the weighted average of the constituent flux measurements for that side of the disk. Major axis separation $< 0$ corresponds to the eastern extent of the disk. Other combinations of NIR filters (e.g. (J-K)) show a predominantly neutral color similar to our (H-K).
\label{fig:rel_color}
}
\end{figure*}

The disk color in isolation from the stellar color can then be analyzed by taking the difference of the relative magnitudes computed above (see Figure \ref{fig:rel_color}), while the east-west asymmetry can be assessed by comparing measurements of opposing sides in a particular bandpass (see Figure \ref{fig:sb_asymmetry}). These results are discussed in Section \ref{sec:disc_asymmetry}).

\subsection{Surface Brightness Results}\label{sec:disc_asymmetry}

\textbf{Disk Color} -- 
Though the nearly unprecedented field of view probed by our CHARIS observations precludes numerical comparisons of photometry with most prior studies, more quantitative comparisons can be made. Our red (STIS/50CCD - BB) color and neutral NIR color measured for the disk (Figure \ref{fig:rel_color}) appear generally consistent with prior literature that diagnosed the inner disk region in optical and NIR. e.g. combining the original discovery observations of \citet{Kalas2007} with new HST/NICMOS imagery, \citet{Debes2008} showed the disk's optical-NIR color becoming redder toward smaller separations.

Numerical simulations in \citet{Boccaletti2003} model disk colors for infrared bandpasses as a function of the dust size distribution's minimum grain size ($\rm{a_{min}}$) and porosity (P). With P$=0$, for 1.6 $\mu m$ ($\sim $ H) versus 2.2 $\mu m$ ($\sim $ K), they show: 
\begin{enumerate}
    \item a blue color for $\rm{a_{min}} \lesssim 0.25$ $\mu m$
    \item a red color for $0.25$ $\mu m \lesssim \rm{a_{min}} \lesssim 2$ $\mu m$
    \item a neutral color for $\rm{a_{min}} \gtrsim 2$ $\mu m$ (and briefly for $\rm{a_{min}} \sim 0.25$ $\mu m$, as the color changes from blue to red)
\end{enumerate}
\citet{Rodigas2012} found a predominantly gray ($K_s-L^{\prime}$) color ($2.1$ and $3.8$ $\mu m$ respectively) across the disk from 1\farcs1 to 1\farcs45. Comparing this result with grain-color models, they suggest a distribution comprised of grains from $\sim 3-10$ $\mu m$. The results of \citet{Boccaletti2003} show that a minimum grain size of $\sim 3-10$ $\mu m$ should also produce a neutral color for (H-K), consistent with our measurements (Figure \ref{fig:rel_color}) and the suggestions of \citet{Rodigas2012}.

However, by comparing measurements in the STIS/50CCD bandpass with our CHARIS broadband measurements, we find a definitively red color throughout the region of overlap ($0\farcs4 - 1\farcs0$). While a wide range of minimum grain size values can produce a neutral color, a much smaller range result in a strong red color. Given that the redder filters analyzed in \citet{Boccaletti2003} predict no significantly red colors for minimum grain sizes larger than $\sim 1$ $\mu m$, the measurement of a very red (STIS/50CCD $-$ CHARIS broadband) color suggests a smaller minimum grain size, $\sim 0.25-1.0$ $\mu m$, is needed to simultaneously produce the red (STIS/50CCD $-$ CHARIS broadband) and gray IR colors that we observe.

\citet{Rodigas2012} compute a blow-out size, $\rm{a_{BO}}$, of $\sim 1-3$ $\mu m$ for HD 15115, indicating that grains of the minimum grain size that we estimate above would likely be expelled from the system. While a larger porosity would increase the estimated minimum grain size, with $\rm{a_{min}}$ $\simprop (1-P)^{-1}$ \citep{Boccaletti2003}, it should also increase the blowout size by a comparable factor, with $\rm{a_{BO}}$ $\simprop (1-P)^{-1}$ \citep{Arnold2019}. A minimum grain size below the theoretical blow-out size can be explained in a number of ways. \citet{Hughes2018} note that this phenomenon is commonly observed and suggest that it is likely the result of a change in grain collision physics near the limit of small grains.  Alternatively, this could manifest if some mechanism for continually replenishing smaller grains is present, such as planetesimal collisions (e.g. \citealt{Hahn2010}).

\textbf{Disk Asymmetry} -- 
Numerous mechanisms have been proposed previously to explain the observed flux asymmetry of HD 15115's ring-like disk and extended halo. In the debris disk's discovery paper, \citet{Kalas2007} suggested the possibility of a past encounter with nearby star HD 12545 perturbing planetesimal orbits to cause the asymmetry. However, \citet{MacGregor2019} points out that the spatial motion of the two objects makes this encounter unlikely. \citet{Debes2009} explored the possibility of disk sculpting through interaction with the interstellar medium (ISM) to explain the bluer western color observed at large separations as well as the observed east-west brightness asymmetry. Since HD 15115's motion lies primarily in the direction of its apparently truncated eastern extent, pressure from clumps of ISM gas might redistribute dust from the eastern side to the western side (assuming motion of the ISM gas itself is favorable for this scenario); with smaller grains being more susceptible to this mechanism, this should result in both a bluer and brighter western disk. The substantial asymmetries in the outer disk halo uniquely revealed by \citet{Schneider2014} STIS imaging provide further evidence that the outer disk and halo are being perturbed. However, using the equations and parameters provided in \citet{Debes2009} gives an approximate `deflection radius' (the stellocentric radius beyond which dust grains are likely to be significantly affected by the interaction) of $100-200$ au. Given CHARIS's $\sim 5-50$ au field of view, it seems unlikely that ISM interactions could be responsible for the asymmetry we observe. Moreover, this interaction should preferentially redistribute smaller grains to the western side-- which is not supported by our disk color and surface brightness asymmetry measurements (See Figures \ref{fig:sb_asymmetry} \& \ref{fig:rel_color}); the fact that we measure a similar overall asymmetry in the STIS and CHARIS broadband data within the CHARIS field of view ($\Delta m \sim 0.6$ and $0.4$ mags respectively) suggests that the phenomenon at work changes the overall dust density between the east and west, without significantly affecting the shape of the grain size distribution.

The results of \citet{Mazoyer2014} showed that while the eastern and western extents are significantly asymmetrical in brightness, the system features a symmetrical ring. This casts doubt on explanations of the brightness asymmetry which would necessitate an observable geometric asymmetry. More recently, \citet{Sai2015} reported an eccentricity for HD 15115's disk of $e=0.06$, which could contribute to the asymmetry we observe. By itself, this eccentricity does not appear capable of producing an asymmetry of the observed size, with limited testing showing an induced east-west asymmetry of $\lesssim 10\%$. However, beyond asymmetry resulting directly from the eccentricity, \citet{Hahn2010} notes that such a system may manifest with asymmetric dust distributions as a result of the difference in orbital velocity between apsides effectively enforcing differing ejection criteria. It is unclear if this mechanism would be capable of producing asymmetry of the observed magnitude.

A number of studies have proposed the possibility of asymmetry resulting from dynamical interactions with an embedded planet -- both for HD 15115 (e.g. \citealt{Sai2015}) and for similar nearly edge-on systems (e.g. HD 111520, \citealt{Draper2016}). \citet{Sai2015} suggests the possibility of planetsimals being trapped in the Lagrange points of an embedded planetary companion. Such an embedded planet might also induce other disk structures: as mentioned in Section \ref{sec:disc_modeling}, complicated disk structures such as spiral arms could be present here but self-obscured by the system's steep inclination. Such a geometry might result in asymmetries similar to those we observe. 

Asymmetries might also be induced by major collisions within the disk \citep{Hahn2010}. The possibility of a minimum grain size below the blow-out size from our color analysis could be explained by this. In contrast to suggestions of planet ``signposts" in similar systems, \citet{Thebault2012} simulates the interactions of debris disks and planets and concludes that for edge-on systems only weak asymmetries will typically result from planet interactions.

\section{Limits on Planets}\label{sec:planet_lims}
Following the procedure described in \citet{Currie2018} for planet forward modeling, we computed $5\sigma$ contrast limits in CHARIS broadband for the planet detection reductions outlined in Section \ref{sec:reduc}. We then mapped these contrasts to planet detection limits using the hot-start, solar metallicity, hybrid cloud, synthetic planet spectra provided by \citet{Spiegel2012}.

\begin{figure*}
\includegraphics[width=\textwidth]{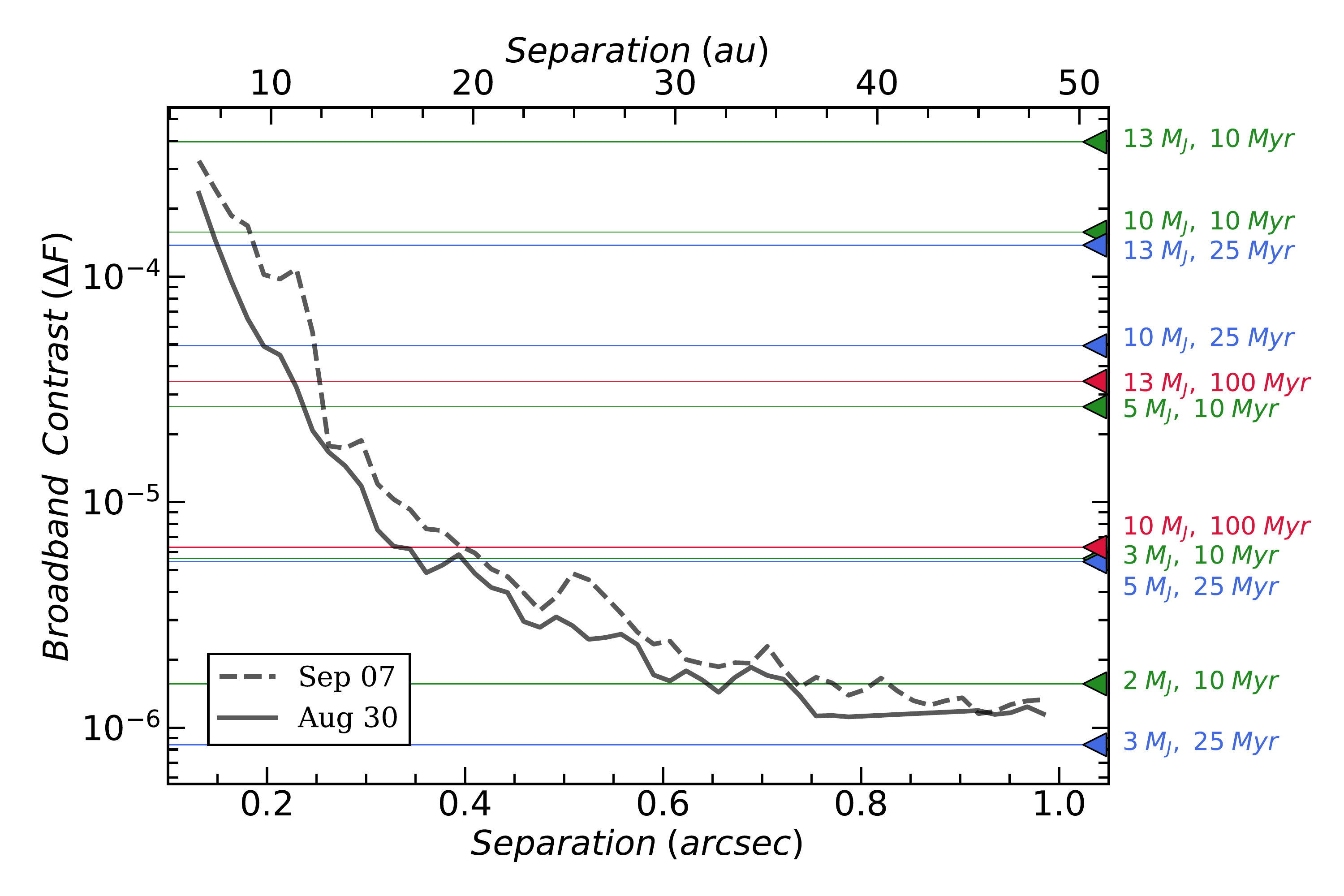}
\caption{Broadband ($1.13-2.39$ $\micron$) contrast curves for planet detection reductions of CHARIS HD 15115 data of August 30 and September 07 outlined in Section \ref{sec:reduc}. $5 \sigma$ contrast is given as a function of stellocentric angular separation (arcsec, lower x-axis) and projected separation (au, upper x-axis). Carets along the right edge of the figure and corresponding colored lines give the planet detection limits for 10 Myr (green), 25 Myr (blue), and 100 Myr (red) planets, based on hot-start, solar-metallicity, hybrid cloud planet evolution models of \citet{Spiegel2012}. The displayed ages are chosen to correspond to various determinations for the HD 15115 system (see Section \ref{sec:planet_lims}).
\label{fig:contrast_curve}
}
\end{figure*}

Model planet spectra corresponding to an array of distinct determinations for the system's age are utilized. These age determinations include: possible membership in TW Hydrae association from Banyan $\Sigma$ (98 \% likelihood; \citealt{Gagne2018}) with age $10\pm3$ Myr \citep{Bell2015}, possible membership in the $\beta$ Pictoris moving group \citep{Moor2006} with age $24\pm3$ Myr \citep{Bell2015}, and various other methods summarized in \citet{Rhee2007} which yield an age of $\sim 100$ Myr. Each planet spectrum was convolved with the filter profile for CHARIS's broadband filter and integrated to determine the photometric bandpass flux. The flux measured for HD 15115 was then converted to an absolute flux (to match the planet spectra) to determine the contrast at which each planet model would manifest. These values are indicated along the right axis of Figure \ref{fig:contrast_curve}. While planet contrast is intrinsically more favorable in the K-band, contrasts achieved are superior for CHARIS broadband imagery to the extent that the broadband offers the strongest constraints on the presence of planets.

The results of this procedure show that our August 30 data reduction is sensitive to $10$ $M_{j}$ companions at the lower and upper suggested ages to separations of $\sim 7.5$ $\rm{au}$ and $\sim 16$ $\rm{au}$ respectively. We note that, given the small mass of the possible companion proposed by \citet{MacGregor2019}, $0.2$ $M_{J}$, we can place no constraints regarding its appearance anywhere within our field of view. On the other hand, for the scenario of a 12 $M_J$ companion at a separation of 45 au discussed in \citet{Sai2015}, we can rule out the planet over the majority of its orbit (e.g. $\sim 93 \%$ of its orbit for an age of 25 Myr).

\section{Conclusions and Future Work}\label{sec:conclusion}
CHARIS imagery of the HD 15115 system has revealed the inner regions of the disk in remarkable detail and probed substantially further than any previous scattered light data (to $\rho \sim 0\farcs2$). This imagery revealed no direct evidence of planetary mass compansions and has allowed for new constraints to be placed on the possibility of a yet-unseen substellar companion in the disk. Combined with the differential evolution algorithm, CHARIS imagery has enabled us to conduct a thorough exploration of the recently proposed inner ring through forward modeling. In doing so, we find a poor fit for a significantly non-coplanar inner ring, but reasonable fits for both a single ring and two rings aligned along our line of sight (either coplanar or manifesting with similar projected semi-minor axes). These data, combined with HST STIS imagery, have allowed for measurement of the disk's color and asymmetry at separations from $0\farcs25$ to $1\farcs0$ and spanning wavelengths from 0.6 $\micron$ to 2.3 $\micron$. These measurements suggest a minimum grain size in the CHARIS field of view of $\lesssim 1.0$ $\micron$, and thus smaller than previous estimates at larger separations.

The CHARIS observations presented here provide the first clear view of the system within $\rho \sim 0\farcs4$. In general, follow-up observations probing this region of the system will better substantiate the results of our analysis. Follow-up observations with CHARIS would enable further constraints to be placed on the presence of inner disk features or companions, as well as gauging the significance of the $\rho \lesssim 0\farcs25$ feature we note in Section \ref{sec:disc_modeling}. The use of CHARIS's new polarimetric integral field spectroscopy mode would allow for measurement of the disk's fractional polarization in CHARIS's field of view, a key diagnostic of the disk's dust properties \citep{Perrin2015}, while also allowing more rigorous assessment of any planet candidates that might be identified. High SNR mid-IR spectra of HD 15115 (e.g. with JWST/MIRI) could better constrain the dust composition within the disk by identifying the signatures of both silicates and non-silicate species using spectral decomposition software \citep{Hughes2018}.

\acknowledgements We thank our referee for providing helpful comments that improved this manuscript.  

The authors wish to acknowledge the very significant cultural role and reverence that the summit of Mauna Kea has always had within the indigenous Hawaiian community.  We are most fortunate to have the opportunity to conduct observations from this mountain.  We wish to acknowledge the critical importance of the current and recent Subaru telescope operators, daycrew, computer support, and office staff employees.  Their expertise, ingenuity, and dedication is indispensable to the continued successful operation of Subaru.  

We acknowledge funding support from the NASA XRP program via grants 80NSSC20K0252 and NNX17AF88G.  T.C. was supported by a NASA Senior Postdoctoral Fellowship.  

Based partially on observations made with the NASA/ESA Hubble Space Telescope, obtained at the Space Telescope Science Institute, which is operated by the Association of Universities for Research in Astronomy, Inc., under NASA contract NAS5-26555. These observations are associated with program \# 12228.

E. A. is supported by MEXT/JSPS KAKENHI grant No. 17K05399.

\bibliography{refs}{}
\bibliographystyle{aasjournal}

\appendix
\section{Scattering Phase Function Comparison}\label{app:phasefuncs}
\begin{figure*}[hb!]
\begin{center}
\includegraphics[width=0.8\textwidth]{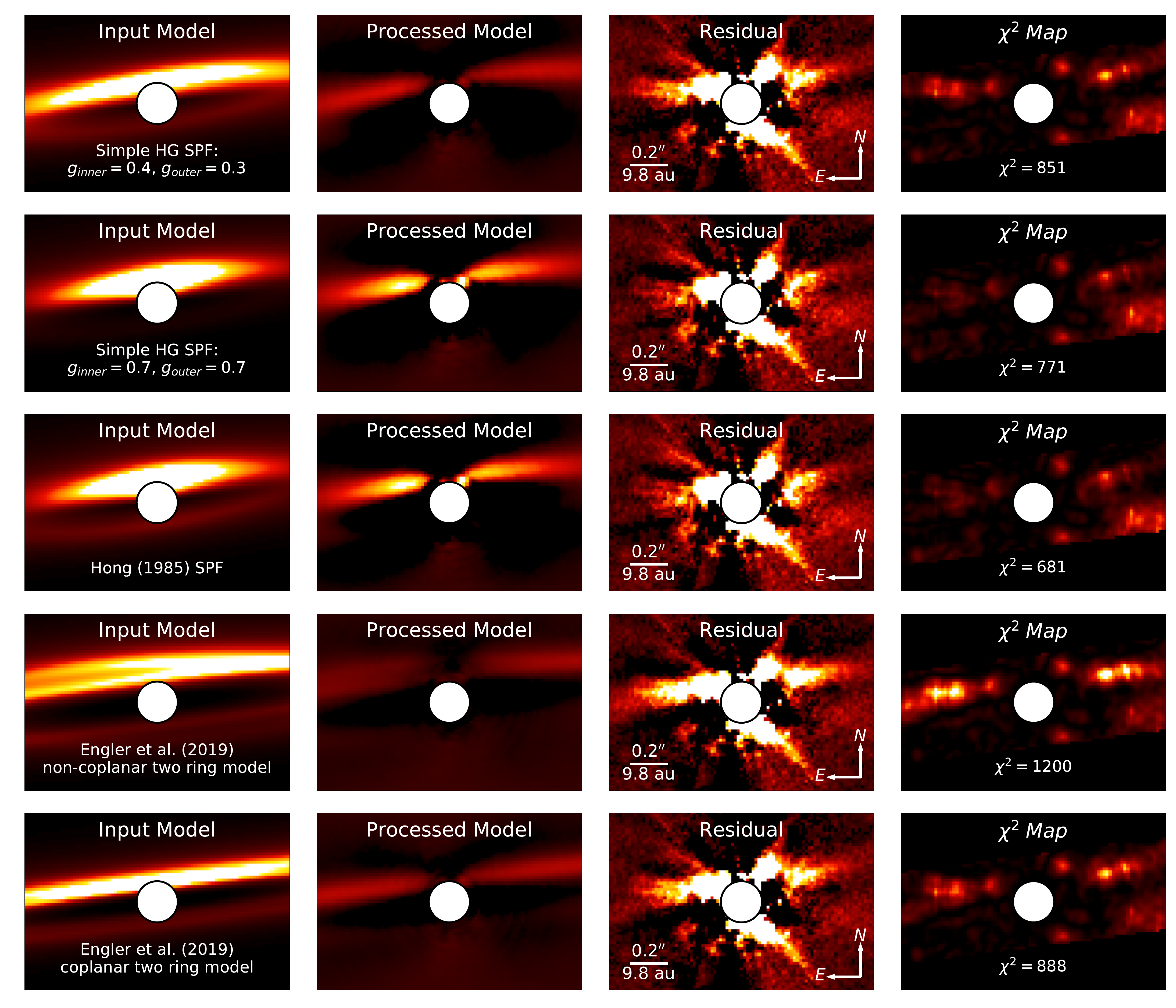}\end{center}
\caption{Each row's images depict (left to right): an initial two ring disk model, the model following attenuation by the forward-modeling procedure for our Aug 30 A-LOCI reduction and rescaling to minimize $\chi^2_\nu$ (see Section \ref{sec:forward_modeling}), the residual after subtracting the model from the data, and the corresponding $\chi^2$ map. The value of $\chi^2$ given in the last panel is for the reduction shown only. In each case, images are shown at the same linear display stretch as the corresponding images in Figure \ref{fig:onecomp_model}. The first three rows show models in which only the phase function changes -- the parameters are otherwise identical and correspond to the best overall two ring disk model identified in Section \ref{sec:forward_modeling}. The model in the first row utilizes the same phase function implemented by \citet{Engler2019} for their two ring models: a simple HG phase function with asymmetry parameter $g = 0.4$ for the inner ring and $g = 0.3$ for the outer. The second row's model changes the asymmetry parameter of both rings to $g = 0.7$, matching the highest weighted term in the Hong phase function. The third row's model utilizes the phase function of \citet{Hong1985}, as adopted for our modeling procedure (see Section \ref{sec:models}). The models in the fourth and fifth row adopt the non-coplanar and coplanar (respectively) best fitting two ring models reported in Section 5.2 of \citet{Engler2019}, which feature the same phase function as the model of the first row, but with differing parameters elsewhere. Note: many of the models utilizing a simple HG phase function appear especially dim in the ``processed model" panels as a result of the rescaling applied at the end of the forward modeling procedure; this is simply the scaling of the model that best minimizes the weighted residuals.
\label{fig:phase_funcs}
}
\end{figure*}

Figure \ref{fig:phase_funcs} shows the results of forward modeling for models of various phase functions for our August 30 data. Overall, a simple HG phase function seems to very poorly describe the brightness profile that we observe.

\section{Python implementation of differential evolution}\label{app:de_code}

Here, we provide a simple Python implementation of the differential evolution algorithm \citep{Storn1997} as described in Section \ref{sec:model_opt}. This code favors readability and simplicity over perfect computational efficiency and has no dependencies besides the NumPy\footnote{\url{https://numpy.org/}} module. In comparison with grid searches, this implementation of DE will result in both a superior fit and orders of magnitude fewer model evaluations for the overwhelming majority of cases.

\begin{lstlisting}
import numpy as np
    
def differential_evolution(objective_fn, converged, bounds, mutation=(0.5,1.), P=0.7, popsize=10):
    ```
    A simple implementation of the differential evolution algorithm using the `best1bin' strategy and allowing a `dithered' mutation constant.
    
    Parameters
    ---------
    objective_fn : callable
        A function that takes the model parameters (1d array of length (popsize*K)) as its argument and returns the value to be minimized (typically chi-squared). This function should: generate the appropriate model from the list of parameters, propagate the model through your forward modeling routine, and then compare the model to your data to determine its fitness. You will probably want to have this function save the input and output models to disk as well.

    converged : callable
        A function that takes the current (normalized) population (2d array of shape (popsize*K, K) for K parameters) and their fit metrics (1d array of length (popsize*K)), returning True if some convergence criteria has been met and fitting should cease and False otherwise.

    bounds : numpy array of shape (K,2) where K is the number of model parameters
        Each entry of `bounds', bounds[i,:], should provide the lower and upper bound for a parameter.

    mutation : float or tuple(float, float), optional
        The mutation constant to utilize. Storn & Price (1997) suggest that values in the range [0.4, 1.0] are typically more favorable. If given as a tuple, the mutation constant is randomly selected each generation from the uniform distribution spanning the two values given.

    P : float, optional
        The crossover probability to utilize. The value of P should be in the range (0,1]. 

    popsize : int, optional
        The number of population members per free parameter to utilize.
        
    Returns
    -------
    : array, float
        The set of best fitting parameters and the associated fitness metric.
    '''
    N,K = bounds.shape[0]*popsize, bounds.shape[0] # number pop. members and parameters
    bmin, brange = bounds[:,0], np.diff(bounds.T, axis=0) # lower lims and range for each param
    x = np.random.rand(N, K) # Generate initial (normed) population array
    fx = np.array([objective_fn(xi) for xi in x*brange+bmin]) # The initial pop's fitness
    indices = np.arange(N) # Define indices corresponding to population members
    while not converged(x,fx): # Loop until converged(x,fx) returns True
        if type(mutation) == tuple: m = np.random.uniform(*mutation) # For dithered m
        else: m = mutation
        xtrial = np.zeros_like(x)
        j = np.argmin(fx) # For best1bin method, j is the index of the best member
        for i in indices:
            k,l = np.random.choice(indices[~np.isin(indices, [i,j])], 2, replace = False)
            xmi = np.clip(x[j] + m*(x[k]-x[l]), 0, 1) # ith mutant vector, clipped to bounds
            xtrial[i] = np.where(np.random.rand(K) < P, xmi, x[i]) # Get trial pop. vector
        fxtrial = np.array([objective_fn(xi) for xi in xtrial*brange+bmin]) # Fitness of trial pop.
        improved = fxtrial < fx # Boolean array indicating which trial members were improvements
        x[improved], fx[improved] = xtrial[improved], fxtrial[improved] # Replace improved members
    return x[np.argmin(fx)]*brange+bmin, np.min(fx) # Return the best params and fitness
\end{lstlisting}

The code as presented can be easily adapted for parallel processing with minor changes to the two lines that evaluate the fitness for a set of model parameters; e.g. using the Joblib module\footnote{\url{https://joblib.readthedocs.io}}, the $4^{\rm{th}}$ line of code in the function could be replaced with (likewise for the $15^{\rm{th}}$ line):

\begin{lstlisting}
    from joblib import Parallel, delayed
    fx = np.array(Parallel()(delayed(objective_fn)(xi) for xi in x*brange+bmin))
\end{lstlisting}
\end{document}